\documentclass[authoryear]{elsarticle}
\journal{Elsevier}
\usepackage[top=20mm, bottom=20mm, left=18mm, right=18mm, showframe=false]{geometry}	
\usepackage{url}
\usepackage{amsfonts,amsmath,amssymb,mathrsfs,amsthm}
\usepackage{graphicx}
\usepackage{dsfont}
\usepackage{natbib}
\usepackage{upgreek}
\usepackage{multirow}

\usepackage[normalem]{ulem}
\usepackage{xcolor}

\newcommand{\pdif}[2]{\frac{\partial #1}{\partial #2}}
\newcommand{\modulus}[1]{\left| \kern.05em #1 \kern.05em \right|}

\newcommand{\captionit}[1]{\caption{\small\textit{#1}}}

\begin{document}
	\begin{frontmatter}
		
		\title{A penalised piecewise-linear model for non-stationary extreme value analysis of peaks over threshold}
		\author[lnc]{Anna Maria Barlow}
		\author[pnr]{Ed Mackay\corref{cor1}}
		\author[lnc]{Emma Eastoe}
		\author[lnc,shl]{Philip Jonathan}
		\address[lnc]{Department of Mathematics and Statistics, Lancaster University LA1 4YF, United Kingdom.}
		\address[pnr]{College of Engineering, Mathematics and Physical Sciences, University of Exeter, Penryn TR10 9FE, United Kingdom.}
		\address[shl]{Shell Research Limited, London SE1 7NA, United Kingdom.}
		\cortext[cor1]{Corresponding author {\tt e.mackay@exeter.ac.uk}}
		\begin{abstract}
        Metocean extremes often vary systematically with covariates such as direction and season. In this work, we present non-stationary models for the size and rate of occurrence of peaks over threshold of metocean variables with respect to one- or two-dimensional covariates. The variation of model parameters with covariate is described using a piecewise-linear function in one or two dimensions defined with respect to pre-specified node locations on the covariate domain. Parameter roughness is regulated to provide optimal predictive performance, assessed using cross-validation, within a penalised likelihood framework for inference. Parameter uncertainty is quantified using bootstrap resampling. The models are used to estimate extremes of storm peak significant wave height with respect to direction and season for a site in the northern North Sea. A covariate representation based on a triangulation of the direction-season domain with six nodes gives good predictive performance. The penalised piecewise-linear framework provides a flexible representation of covariate effects at reasonable computational cost.
		\end{abstract}
		
		
	\end{frontmatter}

\section{Introduction}
\subsection{Background}

Metocean variables such as significant wave height, wind speed and current speed usually show dependence on covariates such as direction, season and water depth (e.g. \citealt{RndEA15}), as well as having complex inter-relationships: e.g. the relationship between wind speed and significant wave height (e.g. \citealt{TowEA17, Mackay2020omae}), or significant wave height and spectral peak period (e.g. \citealt{Hvr87, JntFlnEwn10}), or significant wave height and surge (e.g. \citealt{Ross2018}). Good characterisation of the metocean environment and its extremes demands statistical models which accommodate dependencies of variables on covariates, and dependencies between extremes of variables, in a careful manner.

In this work, we consider a univariate response $Y$ (e.g. storm peak significant wave height, or wind speed) which varies systematically with one or more covariates $\mathbf{X}=(X_1,X_2,...,X_D)$, (e.g. storm direction, season, ...), such that the tail of the conditional distribution $Y|\mathbf{X}=\mathbf{x}$ follows a known parametric form. In a peaks-over-threshold model, we expect $Y|(\mathbf{X}=\mathbf{x},Y>u(\mathbf{x}))$ to follow a generalised Pareto (GP) distribution, provided that the extreme value threshold $u(\mathbf{x})$ is sufficiently large, with shape, scale and threshold parameters $\xi(\mathbf{x})$, $\sigma(\mathbf{x})>0$ and $u(\mathbf{x})$ varying systematically with $\mathbf{x}$, and density
\begin{equation} \label{eq:GP_dens}
        f_{\text{GP}}(y|u(\mathbf{x}),\sigma(\mathbf{x}),\xi(\mathbf{x}))= 
        \begin{cases}
        \dfrac{1}{\sigma(\mathbf{x})} \left[1 + \xi (\mathbf{x}) \left(\dfrac{y-u(\mathbf{x})}{\sigma(\mathbf{x})}\right) \right] _{+}^{-1-1/\xi(\mathbf{x})}, & \xi(\mathbf{x}) \neq 0, \\
        \dfrac{1}{\sigma(\mathbf{x})} \exp \left(- \left(\dfrac{y-u(\mathbf{x})}{\sigma(\mathbf{x})}\right) \right), & \xi(\mathbf{x}) = 0,
        \end{cases}
\end{equation} 
where $[\cdot]_+ = \max\lbrace \cdot,0 \rbrace$. Using a similar approach, we can also model the rate of occurrence of events, dependent on covariates, $\mathbf{X}$, assuming a Poisson model for counts of exceedances.

In more applied work, the effects of covariates have been either ignored historically, or accommodated by performing independent extreme value inferences for subsets of the data corresponding to different parts of the covariate domain. Partition of the sample into smaller samples on sub-domains can result in increased and unquantified bias and uncertainties if the sample on each sub-domain is treated independently, without taking into account the systematic variation of the response with the covariate (e.g. \citealt{Mackay2010}, \citealt{Mackay2020assessment}). A better approach is to allow the parameter of the extreme value model to vary with covariate. Numerous such covariate models have been developed for extreme value analysis, including \cite{CrtChl81}, \cite{DvsSmt90}, \cite{ClsWls94}, \cite{RbsTwn97}, \cite{SctGds00}, \cite{AndCrt01}, \cite{ChvDvs05}, \cite{FwcWls07}, \cite{MndEA08}, \cite{NrtJnt11}, \cite{RndEA15}, \cite{Siguake2017} and \cite{Eastoe2019}. \cite{ZnnEA19a} provides a general description of semi-parametric covariate representations for extreme value analysis of peaks over threshold, including penalised B-splines, Bayesian adaptive regression splines and Voronoi partitions. These all take the form of a linear combination of basis functions on the covariate domain, but differ (a) in the way that basis functions are constructed and modified, and (b) by additional penalisation of the variability (e.g. variance or roughness) of basis coefficients, for a given sample, to improve inference. Inference using these representations is generally computationally complex, requiring Bayesian inference for all but the simplest representations.
 
In marked contrast, \cite{Ross2018} use a penalised piecewise-constant (PPC) model. The motivation for this approach is the provision of a simple, pragmatic but useful representation for covariate effects, both in terms of parameter estimation and uncertainty quantification. In the PPC model, the multi-dimensional covariate domain $\mathcal{D}$ is partitioned into sub-domains or ``bins'' $B_1, B_2, ..., B_M$. In each bin $B_k$, the distribution of $Y|(\mathbf{X}=\mathbf{x},Y>u_k)$, $\mathbf{x} \in B_k$ is assumed to be stationary and follow a GP distribution with density given in Equation~\ref{eq:GP_dens}, with pre-specified extreme value threshold $u(\mathbf{x})=u_k$ and scale $\sigma(\mathbf{x})=\sigma_k$. The shape parameter $\xi$ is assumed common to all bins. The objective of the PPC inference is simultaneous estimation of the scale parameters and common shape parameter for all $M$ bins. This is achieved by maximisation of predictive likelihood for a hold-out sample of data, further penalised by the roughness of the set $\sigma_1, \sigma_2, ..., \sigma_M$. The roughness penalty is selected to maximise the predictive likelihood, evaluated using a cross-validation scheme. Bootstrap resampling is used to quantify uncertainties of estimated parameters and other predictions. The PPC model has been used in a number of applications (e.g. \cite{RssEA19}, \cite{Mackay2020assessment}, \citealt{Guerrero2021}). 

In \cite{Mackay2020assessment}, it was observed that the piecewise-constant assumption did not provide the most natural, parsimonious representation for parameter variation on the covariate domain for some applications of interest, and that a piecewise-linear covariate parameterisation may be more appropriate. These observations motivated initial research into a penalised piecewise-linear (PPL) model outlined in \cite{Mackay2020omae} (using common shape parameter and one-dimensional covariates) and subsequently the full development for arbitrary covariate domains reported here. There is a long history of piecewise linear or segmentation regression, motivated by the assumption of local smooth (and in particular linear) variation of model parameters with covariates (e.g. \citealt{Clv79}, \citealt{YngEA16}). Like PPC, the PPL model is intended to provide a useful compromise between simplicity and flexibility. Like PPC, PPL is simpler to implement and compute than extreme value models incorporating general basis function representations for covariates discussed by \cite{ZnnEA19a}. At the same time, relative to PPC, PPL provides a more flexible and physically-realistic covariate representation at the cost of increased computational complexity.

\subsection{Motivating application} \label{sec:App:Data}
We apply the non-stationary PPL extreme value model to data from the NORA10 hindcast of \cite{RstEA11}, for a location in the northern North Sea. The hindcast provides time-series of significant wave height, (dominant) wave direction and season (defined as day of the year, for a standardised year consisting of 360 days) for three hour sea-states for the period 1957-2010. Throughout, ``direction'' refers to the direction from which a storm travels expressed in degrees clockwise with respect to north. Since significant wave height is serially-correlated between sea states, storm peak significant wave height characteristics are isolated from the hindcast time-series using the procedure described in \cite{EwnJnt08}. Contiguous intervals of significant wave height above a low peak-picking threshold are identified, each interval now assumed to correspond to a storm event. The peak-picking threshold corresponds to a directional-seasonal quantile of significant wave height with specified non-exceedance probability, estimated using quantile regression. The maximum of significant wave height during the storm interval is taken as the storm peak significant wave height (henceforth $H_S$). The values of directional and seasonal covariates at the time of storm peak significant wave height are referred to as storm peak values of those variables. The resulting storm peak sample consists of 5388 values of $H_S$. The seasonal and directional characteristics of the sample are obvious from inspection of the figures in Sections~\ref{sec:App:1D} and \ref{sec:App:2D}. The directional influence of the land shadow of Norway is clear on the interval $(45^\circ, 210^\circ)$. A more gradual seasonal variation is also present. We expect these sources of non-stationarity to be adequately characterised in the estimated PPL model. Data from the same location have been considered in \cite{RndEA15a} and \cite{KnzEA21}. 

\subsection{Objectives and layout}
The objectives of the current work are to extend the PPL model of \cite{Mackay2020omae} to incorporate non-stationary shape parameter and multi-dimensional covariates, and to provide software for applications in one and two dimensions. Further, we seek to demonstrate the usefulness of the PPL model in application to non-stationary extreme value analysis of storm peak significant wave height with directional-seasonal covariates for a location in the northern North Sea. The layout of the paper is as follows. Section~\ref{sec:MdlFrm} describes the PPL model formulation in detail, and Section~\ref{sec:Inf_GPPPL} the approach to inference. Sections~\ref{sec:App:1D} and \ref{sec:App:2D} discuss the North Sea application for covariates in one and two dimensions. Finally, Section~\ref{sec:Dsc} provides discussion and conclusions.

\section{Model and preliminaries} \label{sec:MdlFrm}
This section introduces the extreme value model with piecewise-linear covariate representation. It also provides a summary of the approaches used to estimate the density $f_\mathbf{X}$ of covariates, and the non-stationary extreme value threshold $u$. A key motivation for the modelling approach is that covariate effects are more easily identified, and are more influential, in some aspects of the analysis than others (e.g. \citealt{AndCrt01}). In particular, covariate dependence of $f_\mathbf{X}$ and $u$ is relatively-easily identified. It is therefore reasonable to use more flexible non-stationary non-parametric methods to capture these effects. Covariate dependence of GP scale $\sigma$ is also identifiable in general, but typically not to the same extent as that of $f_\mathbf{X}$ and $u$, suggesting a simpler covariate representation for $\sigma$ may be appropriate. Quantifying the covariate dependence of GP shape $\xi$ is most problematic; we anticipate that the piecewise-linear representation, or even a stationary estimate, will be suitable in general.

\subsection{Formulation} \label{sec:MdlFrm:Jnt}
The joint density of covariates $\mathbf{X}$ and response $Y$ can be written as $f_{\mathbf{X},Y}(\mathbf{x},y) =$ $f_{\mathbf{X}}(\mathbf{x}) f_{Y|\mathbf{X}}(y|\mathbf{x})$, where $f_{\mathbf{X}}(\mathbf{x})$ is the density of covariates. We assume that the tail of $f_{Y|\mathbf{X}}(y|\mathbf{x})$ converges to a GP distribution. Our model for exceedances of high non-stationary threshold  $u(\mathbf{x})$, can therefore be written $f_{Y|\mathbf{X}}(y|\mathbf{x}) =\zeta(\mathbf{x}) f_{\text{GP}}\left(y\,|\, u(\mathbf{x}), \sigma(\mathbf{x}), \xi(\mathbf{x})\right)$, $y>u(\mathbf{x})$, where $\zeta(\mathbf{x})=\Pr\left(Y>u(\mathbf{x})\, |\, \mathbf{X}=\mathbf{x} \right)$ is the threshold exceedance probability,  and $f_{\text{GP}}$ is the density of the GP distribution with shape, scale and threshold parameters $\xi(\mathbf{x})$, $\sigma(\mathbf{x})$ and $u(\mathbf{x})$, defined in Equation~\ref{eq:GP_dens}. Hence, for $y > u(\mathbf{x})$, our model for the joint density is
\begin{equation} \label{eq:XY_taildensity}
    f_{\mathbf{X},Y}(\mathbf{x},y) = f_{\mathbf{X}}(\mathbf{x}) \zeta(\mathbf{x}) f_{\text{GP}}\left(y\,|\, u(\mathbf{x}), \sigma(\mathbf{x}), \xi(\mathbf{x})\right), \quad y > u(\mathbf{x}).
\end{equation}
Inference requires estimation of the covariate density $f_{\mathbf{X}}(\mathbf{x})$, a non-stationary threshold $u(\mathbf{x})$ (typically corresponding to a constant values of $\zeta$), and non-stationary GP model with parameters $\sigma(\mathbf{x})$ and $\xi(\mathbf{x})$. For $y \leq u(\mathbf{x})$, $f_{\mathbf{X},Y}(\mathbf{x},y)$ can be estimated empirically, e.g. using kernel density estimation. Moreover, although this is not a requirement of the model in general, in the current work, we assume that covariates $\mathbf{X}$ are periodic and therefore never extreme, so that kernel density estimation is also appropriate to estimate $f_{\mathbf{X}}(\mathbf{x})$. Threshold $u(\mathbf{x})$ will be estimated as a local quantile of $Y$ on the covariate domain. It then remains to estimate the GP model, with parameters $\sigma(\mathbf{x})$ and $\xi(\mathbf{x})$ which vary systematically as a function of covariate, taking (penalised) piecewise linear forms. 

\subsection{Piecewise-linear covariate representation} \label{sec:interp}

The piecewise-linear representation for functions $\sigma(\mathbf{x})$ and $\xi(\mathbf{x})$ of covariate $\mathbf{x} \in \mathcal{D}$ is defined as follows. We first locate $K$ nodes $\mathbf{n}_k$, $k=1,2,...,K$ on the covariate domain $\mathcal{D}$, with coordinates $\mathbf{n}_k=(n_{k,1},...,n_{k,D})$. Next we define a triangulation on $\mathcal{D}$, with nodes as vertices, which partitions $\mathcal{D}$ into $M$ covariate bins $B_m$, $m=1,2,...,M$, each of which is a $D$-simplex. For problems with a single covariate ($D=1$), each $B_m$ is a line segment; for $D=2,3$, $B_m$ is a triangle and tetrahedron respectively. GP model parameter values are specified at the vertices of the triangulation only, and are assumed to vary linearly within each bin $B_m$. The relationship between $M$ and $K$ depends on $D$, and the choice of (non-unique) triangulation of nodes. The assumed periodicity of covariates means that some care is needed when interpolating within bins near the boundary of $\mathcal{D}$.

The vertices of bin $B_m$ are indexed using the indices of the nodes. For bin $B_m$, define the index vector $\mathbf{t}_m=(t_{m,1},\hdots,t_{m,D+1})$, where $t_{m,j}\in\lbrace1,...,K\rbrace$ and $t_{m,i}\neq t_{m,j}$ for $i\neq j$, so that $\lbrace \mathbf{n}_{t_{m,1}},..., \mathbf{n}_{t_{m,D+1}}\rbrace$ is the set of nodes defining the vertices of $B_m$. For an arbitrary function $\psi:\mathbb{R}^D\to\mathbb{R}$, whose values are specified at the nodes, the linear interpolant of $\psi$ for a point $\mathbf{x}=(x_1,x_2,...,x_D)\in B_m$ is given by $\psi(\mathbf{x}) = \tilde{\mathbf{x}} \boldsymbol{\beta}_m$, where $\tilde{\mathbf{x}}=(1,x_1,x_2,...,x_D)$, and $\boldsymbol{\beta}_m = (\beta_{m,0},...,\beta_{m,D})^T$ is the solution of ${A}_m \boldsymbol{\beta}_m = \mathbf{c}_m$, where 
\begin{equation*}
	{A}_m = 
	\begin{pmatrix}
		1 & \mathbf{n}_{t_{m,1}}\\
		\vdots & \vdots  \\
		1 & \mathbf{n}_{t_{m,D+1}} 
	\end{pmatrix}
	= 
	\begin{pmatrix}
		1 & n_{t_{m,1},1} & \hdots n_{t_{m,1},D}\\
			\vdots & \vdots  \\
		1 & n_{t_{m,D+1},1} & \hdots n_{t_{m,D+1},D}
	\end{pmatrix}
\end{equation*}
and vector $\mathbf{c}_m=\left(\psi\left(\mathbf{n}_{t_{m,1}}\right), \hdots, \psi\left(\mathbf{n}_{t_{m,D+1}}\right) \right)^T$ stores the current values of $\psi$ at the nodes $\mathbf{t}_m$. Provided that $D$ is not too large, it is straightforward to calculate ${A}_m^{-1}$ for each bin $B_m$, and save it in memory for repeated use, making iterative estimation of a piecewise linear model computationally efficient.

\subsection{Estimation of covariate density and extreme value threshold}  \label{sec:Inf_CvrDns_Thr}
Consider a sample $\mathcal{S} = \lbrace (\mathbf{x}_1,y_1),\hdots, (\mathbf{x}_N,y_N)\rbrace$ of $N$ conditionally-independent observations, to be used to estimate the model in Equation~\ref{eq:XY_taildensity}. This section describes inference for covariate density $f_\mathbf{X}(\mathbf{x})$ and threshold $u(\mathbf{x})$, with penalised piecewise-linear inference for the GP model described in Section~\ref{sec:Inf_GPPPL}.

\subsection*{Covariate density} \label{sec:cov_dens}
The covariate density $f_{\mathbf{X}}(\mathbf{x})$ is estimated using kernel density (KD) estimation, for points on a regular grid in $\mathcal{D}$. At location $\mathbf{x}$, the KD estimate is
\begin{equation*}
    \hat{f}_{\mathbf{X}}(\mathbf{x}) = \frac{1}{N} \sum_{i=1}^N \prod_{d=1}^D \phi\left(\frac{x_d-x_{i,d}}{w}\right)
\end{equation*}
where $\phi$ is the standard normal density, and common bandwidth $w$ is user-specified to give a reasonable compromise between smoothness and resolution. We note that many more sophisticated models for non-parametric density estimation are available, but this approach appeared adequate for the application discussed in Sections~\ref{sec:App:1D} and \ref{sec:App:2D}.

\subsection*{Threshold estimation} \label{sec:thresh}
In the current work, the threshold exceedance probability $\zeta$ is set prior to analysis, hence threshold estimation reduces to estimation of conditional quantiles $\lbrace u(\mathbf{x}) : \Pr\left(Y>u(\mathbf{x})\, |\, \mathbf{X}=\mathbf{x} \right) = \zeta \rbrace$. We choose to use a two-step process comprised of (1) local quantile estimation and (2) subsequent quantile smoothing for this purpose, but again note that many alternative approaches would be suitable. The local quantile is again estimated on a regular grid in $\mathcal{D}$. For each location $\mathbf{x}$, we find the $C$ nearest observations in $\mathcal{S}$, calculate the empirical quantile corresponding to exceedance probability $\zeta$, and smooth using a Gaussian kernel. The value of $C$ and the bandwidth of the smoothing kernel are modelling choices. 

\section{Estimation of penalised piecewise-linear generalised Pareto model}  \label{sec:Inf_GPPPL}
This section describes maximum roughness-penalised likelihood estimation of GP shape $\xi(\mathbf{x})$ and scale $\sigma(\mathbf{x})$, taking piecewise-linear forms for $\mathbf{x} \in \mathcal{D}$. A critical precursor to successful inference is reasonable node placement on $\mathcal{D}$, discussed in Section~\ref{sec:node}. Given nodes, the maximum likelihood inference is discussed in Section~\ref{sec:likelihood}. The use of cross-validation for optimal choice of roughness penalties is discussed in Section~\ref{sec:crossval}.

\subsection{Node placement and triangulation} \label{sec:node}
We seek to locate nodes on $\mathcal{D}$ such that the piecewise-linear representation is able to describe variation of GP parameters well. Intuitively, this suggests that nodes be located where the gradient of parameters changes sharply with respect to covariate. This is particularly important when there are only a few nodes. To guide selection of node positions, we calculate initial estimates of $\sigma(\mathbf{x})$ and $\xi(\mathbf{x})$ on a regular grid on $\mathcal{D}$, in a similar manner to threshold estimation outlined in Section~\ref{sec:Inf_CvrDns_Thr}. For each $\mathbf{x}$, we find the $C$ nearest threshold exceedances of $u(\mathbf{x})$, and use these to calculate (locally-stationary) moment estimators $\hat{\xi}(\mathbf{x})=(1-(m(\mathbf{x}))^2/v(\mathbf{x}))/2$ and $\hat{\sigma}(\mathbf{x})=m(\mathbf{x}) (1-\hat{\xi}(\mathbf{x}))$, where $m(\mathbf{x})$ and $v(\mathbf{x})$ are the mean and variance of the $C$ nearest threshold exceedances, $z_{j(\mathbf{x})} = y_{j(\mathbf{x})} - u(\mathbf{x}_{j(\mathbf{x})})$, $j(\mathbf{x})=1,...,C$, where $j(\mathbf{x})$ indicates the index of the $j^{\text{th}}$ closest observation to $\mathbf{x}$. If $\hat{\xi}(\mathbf{x})<0$ and $z_{max}(\mathbf{x}) > - \hat{\sigma} (\mathbf{x}) / \hat{\xi} (\mathbf{x})$, where $z_{max}(\mathbf{x}) = \max\lbrace z_{j(\mathbf{x})}: j(\mathbf{x})=1,...,C \rbrace$, then we set $\hat{\xi} (\mathbf{x}) = - \hat{\sigma} (\mathbf{x}) / z_{max}(\mathbf{x})$, since for a negative shape parameter the upper end point of the GP distribution is $-\sigma/\xi$. If $C$ is large, then the true values of the parameters may have a large variation over the range of $\mathbf{x}$ from which the sample is drawn, leading to bias in the estimates. Alternatively, if $C$ is small, then the estimates will be subject to higher variance. Nevertheless, it is assumed that it is possible to choose $C$ such that these initial estimates are adequate to identify gross features of $\sigma(\mathbf{x})$ and $\xi(\mathbf{x})$ on $\mathcal{D}$. Once node locations have been specified, aided by the local estimates of $\sigma$ and $\xi$, we triangulate the covariate domain. For $D=1$, the partition achieved is unique, with bins corresponding to intervals between adjacent nodes. For $D>1$, the triangulation is not unique, and different options are possible as discussed in Section~\ref{sec:App:2D}; clearly, the choice of triangulation also affects model performance.

\subsection{Parameter estimation for given roughness coefficients}
\label{sec:likelihood}
Estimates of the GP scale and shape parameters at the nodes are found by maximising the sample likelihood, penalised for parameter roughness on $\mathcal{D}$. Optimal roughness penalties are estimated by cross-validation, described in Section~\ref{sec:crossval}. 

With parameters $\theta = (\sigma(\mathbf{n}_1), \hdots, \sigma(\mathbf{n}_K), \xi(\mathbf{n}_1), \hdots, \xi(\mathbf{n}_K))$ to be estimated, the sample likelihood for $\mathcal{S}$ under the PPL model is $\mathcal{L}(\theta \, | \,\mathcal{S}) = \prod_{i\in\mathcal{I}} f_{\text{GP}}(y_i\,|\,u(\mathbf{x}_i),\sigma(\mathbf{x}_i),\xi(\mathbf{x}_i))$, where $\mathcal{I}$ is the index set of exceedances $\lbrace i : y_i > u(\mathbf{x}_i) \rbrace$, and for $\mathbf{x} \in B_m$, parameters $\sigma(\mathbf{x})$ and $\xi(\mathbf{x})$ take piecewise-linear representations parameterised in terms of the corresponding bin nodes. The sample negative log-likelihood is denoted $\ell(\theta \, | \, \mathcal{S}) = -\log(\mathcal{L}(\theta \, | \, \mathcal{S}))$.

Roughness penalisation is used to smooth the variation of parameters on $\mathcal{D}$ for optimal predictive performance.  This is quantified using the gradient of the parameter function in each covariate dimension, in each bin. The partial derivatives of function $\psi(\mathbf{x})$ for a point $\mathbf{x}\in B_m$, are given by the interpolation coefficients, $\boldsymbol{\beta}_m$, defined in Section \ref{sec:interp}. That is, $[\partial \psi(\mathbf{x})/\partial x_d]_{\mathbf{x} \in B_m} = \beta_{m,d}$. The penalised negative log-likelihood becomes
\begin{equation} \label{eq:PNLL}
    \ell^*(\theta\,|\,\boldsymbol{\lambda},\mathcal{S}) =  \ell(\theta \,|\,\mathcal{S}) + \sum_{d=1}^{D}\sum_{m=1}^{M} \left( \lambda_{\sigma, d}\,\modulus{\pdif{\sigma(\mathbf{x})}{x_d}\Bigr|_{\mathbf{x} \in B_m}} + \lambda_{\xi, d}\,\modulus{\pdif{\xi(\mathbf{x})}{x_d}\Bigr|_{\mathbf{x} \in B_m}} \right)
\end{equation}
where $\boldsymbol{\lambda} = (\lambda_{\sigma, 1}, \hdots, \lambda_{\sigma, D}, \lambda_{\xi, 1}, \hdots, \lambda_{\xi, D})$ is the vector of roughness penalties. For given $\boldsymbol{\lambda}$, constrained non-linear optimisation is used to minimise $\ell^*(\theta\,|\,\lambda,\mathcal{S})$. In the current work, we use the MATLAB function \textit{fmincon}, under the constraint $-0.5\leq \xi(\mathbf{x}) < 0$, reasonable for environmental variables (with $\xi=-0.5$ corresponding to linear decrease in GP density with $y$, and $\xi<0$ ensuring a finite upper bound for the distribution of $Y$). A large value of $\lambda$ for some parameter-covariate pair imposes a large penalty on the corresponding parameter gradient, and hence drives smoother solutions for that parameter-covariate pair. The starting solution for the optimisation is found using a Voronoi partition of $\mathcal{D}$, with bins corresponding to the set of points closest to each node. Independent GP fits, constrained such that $-0.5\leq \xi(\mathbf{x}) < 0$, are then made per Voronoi bin, and the parameter estimates are used as a starting solution for the PPL estimation. In some applications, we may wish to assume that the GP shape parameter does not vary with covariate. In this case the model form can be simplified in the obvious way.

\subsection{Estimation of optimal roughness coefficients} 
\label{sec:crossval}
Section~\ref{sec:likelihood} describes how the parameter set $\theta$ is estimated, given roughness coefficients $\boldsymbol{\lambda}$. The value of $\boldsymbol{\lambda}$ is selected to maximise the predictive performance of the model using a $G$-fold cross-validation scheme. Cross-validation groups $\mathcal{S}_g$, $g=1,2,...,G$ are formed by random partition of the sample $\mathcal{S}$ into $G$ groups of approximately equal size. 

For each of a large number $L$ of plausible choices for $\boldsymbol{\lambda}$, and cross-validation group $\mathcal{S}_g$, we estimate the GP model using $\mathcal{S} \setminus \mathcal{S}_g$, finding parameter vector $\theta_g(\boldsymbol{\lambda})$ that minimises the penalised negative log-likelihood $\ell^*(\theta\,|\,\boldsymbol{\lambda},\mathcal{S} \setminus \mathcal{S}_g)$. We then calculate the corresponding unpenalised negative log predictive likelihood $\ell(\theta_g(\boldsymbol{\lambda}) \, | \, \mathcal{S}_g)$ for the excluded group $\mathcal{S}_g$, using optimal parameter vector $\theta_g(\boldsymbol{\lambda})$. The predictive performance of the model for penalty vector $\boldsymbol{\lambda}$ is accumulated as $\mathcal{P}(\boldsymbol{\lambda}) = \sum_{g=1}^{G} \ell(\theta_g(\boldsymbol{\lambda}) \, | \, \mathcal{S}_g)$. The optimal value of $\boldsymbol{\lambda}$ is that which minimises $\mathcal{P}(\boldsymbol{\lambda})$.

Since random partitioning into groups leads to uncertainty in the predictive performance $\mathcal{P}(\boldsymbol{\lambda})$, the procedure is repeated $R$ times, with different replicates of random partitions of $\mathcal{S}$. The overall predictive performance of the model is quantified using $\bar{\mathcal{P}}(\boldsymbol{\lambda})=\frac{1}{R} \sum_{r=1}^R \mathcal{P}_r(\boldsymbol{\lambda})$, where $\mathcal{P}_r(\boldsymbol{\lambda})$ is the predictive performance estimated for replicate $r$. The uncertainty in $\bar{\mathcal{P}}(\boldsymbol{\lambda})$ is calculated using a jackknife procedure. In outline, $R$ different jackknife estimates of $\bar{\mathcal{P}}(\boldsymbol{\lambda})$ are calculated, using all $R$ estimates $\mathcal{P}_{r'}(\boldsymbol{\lambda})$, $r'=1,2,...,R$ except for the $r^\text{th}$ estimate, $r=1,2,...,R$. The uncertainty $U(\bar{\mathcal{P}}(\boldsymbol{\lambda}))$ in the mean predictive performance $\bar{\mathcal{P}}(\boldsymbol{\lambda})$ is quantified as the range of the $R$ jackknife estimates. The optimal choice of $\boldsymbol{\lambda}$ can be adapted to accommodate uncertainty due to the cross-validation strategy. In practice, we might select $\boldsymbol{\lambda}^*$ such that $\bar{\mathcal{P}}(\boldsymbol{\lambda}^*) \leq \bar{\mathcal{P}}(\boldsymbol{\lambda}^\circ) + U(\bar{\mathcal{P}}(\boldsymbol{\lambda}^\circ))$, where $\boldsymbol{\lambda}^\circ$ is the value that minimises $\bar{\mathcal{P}}$, and the values of the components of $\boldsymbol{\lambda}^*$ are larger than those of $\boldsymbol{\lambda}^\circ$. Hence, the estimated model using $\boldsymbol{\lambda}^*$ will be stiffer and therefore more parsimonious than that using $\boldsymbol{\lambda}^\circ$, whilst constraining the degradation in predictive performance to an acceptable level.

\subsection{Computational complexity of predictive inference} \label{sec:App:CmpCmp}
The inference procedure involves iterative accumulation of predictive performance measure $\bar{\mathcal{P}}$ over $L$ roughness coefficients, $G$ cross-validation groups, and $R$ repetitions of the cross-validation partition. For a $D$-dimensional covariate domain $\mathcal{D}$, exploring a reasonable number of values for each of the $2 \times D$ components of the vector $\boldsymbol{\lambda}$ is necessary for reliable performance; searching over $S$ values for $2D$ roughness coefficients requires $L = S^{2D}$ iterations. Computational complexity for the full analysis therefore scales as $G \times R \times S^{2D}$. Assuming that GP shape is stationary on $\mathcal{D}$ reduces computational complexity to $G \times R \times S^{D}$.

In light of this, to restrict computational time to reasonable values for physically-reasonable model complexity, we restrict our interest to the following archetypes. Case A is the computationally-simplest analysis undertaken, for which the GP shape parameter is assumed stationary on $\mathcal{D}$, and a common roughness coefficient is assumed for GP scale with respect to each covariate dimension (i.e. $\lambda_{\sigma,1} = \lambda_{\sigma,2} := \lambda_{\sigma}$ for a $D=2$ directional-seasonal problem). Case B assumes that both GP shape and scale are non-stationary on $\mathcal{D}$, with common roughness coefficients for each with respect to each covariate dimension (i.e. $\lambda_{\sigma,1} = \lambda_{\sigma,2}:=\lambda_{\sigma}$, $\lambda_{\xi,1} = \lambda_{\xi,2}:= \lambda_{\xi}$ for a $D=2$ directional-seasonal problem). Finally, Case C again assumes that GP shape is stationary on $\mathcal{D}$, but estimates separate roughness coefficients for GP scale and each covariate dimension (i.e. $\lambda_{\sigma,1}$ and $\lambda_{\sigma,2}$ for a $D=2$ directional-seasonal problem). Note that for $D=1$, Cases A and C coincide.

After some experimentation for the North Sea application introduced in Section~\ref{sec:App:Data}, and detailed in Sections~\ref{sec:App:1D} and \ref{sec:App:2D} below, we choose to adopt $G=5$, $R=5$ and $S=10$ as reasonable values for the numbers of cross-validation groups, partition repetitions and (component) roughness coefficients respectively. The $S$ values of component roughness coefficients considered are equal logarithmic increments on the interval $[10^{-1},10^5]$, ensuring reasonable coverage from very flexible to very stiff covariate representations. 

\section{Application with one-dimensional covariate} \label{sec:App:1D}
Here we discuss the estimation of one-dimensional directional and seasonal PPL extreme value models for the data described in Section~\ref{sec:App:Data}, following the procedure outlined in Section~\ref{sec:MdlFrm:Jnt}. Two-dimensional directional-seasonal models are presented in Section~\ref{sec:App:2D}.

\begin{figure}
    \centering
    \includegraphics[scale=0.5]{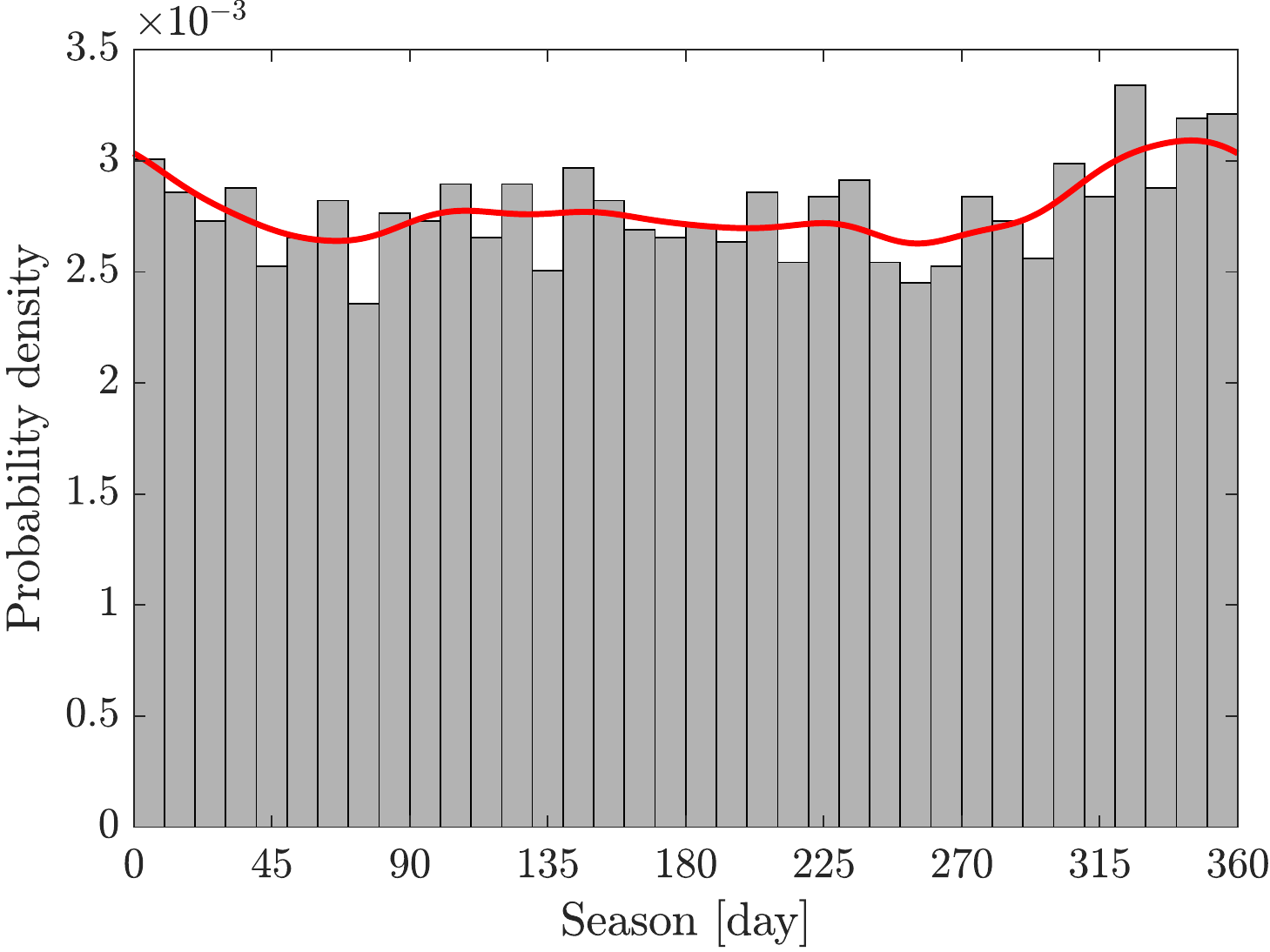} 
    \includegraphics[scale=0.5]{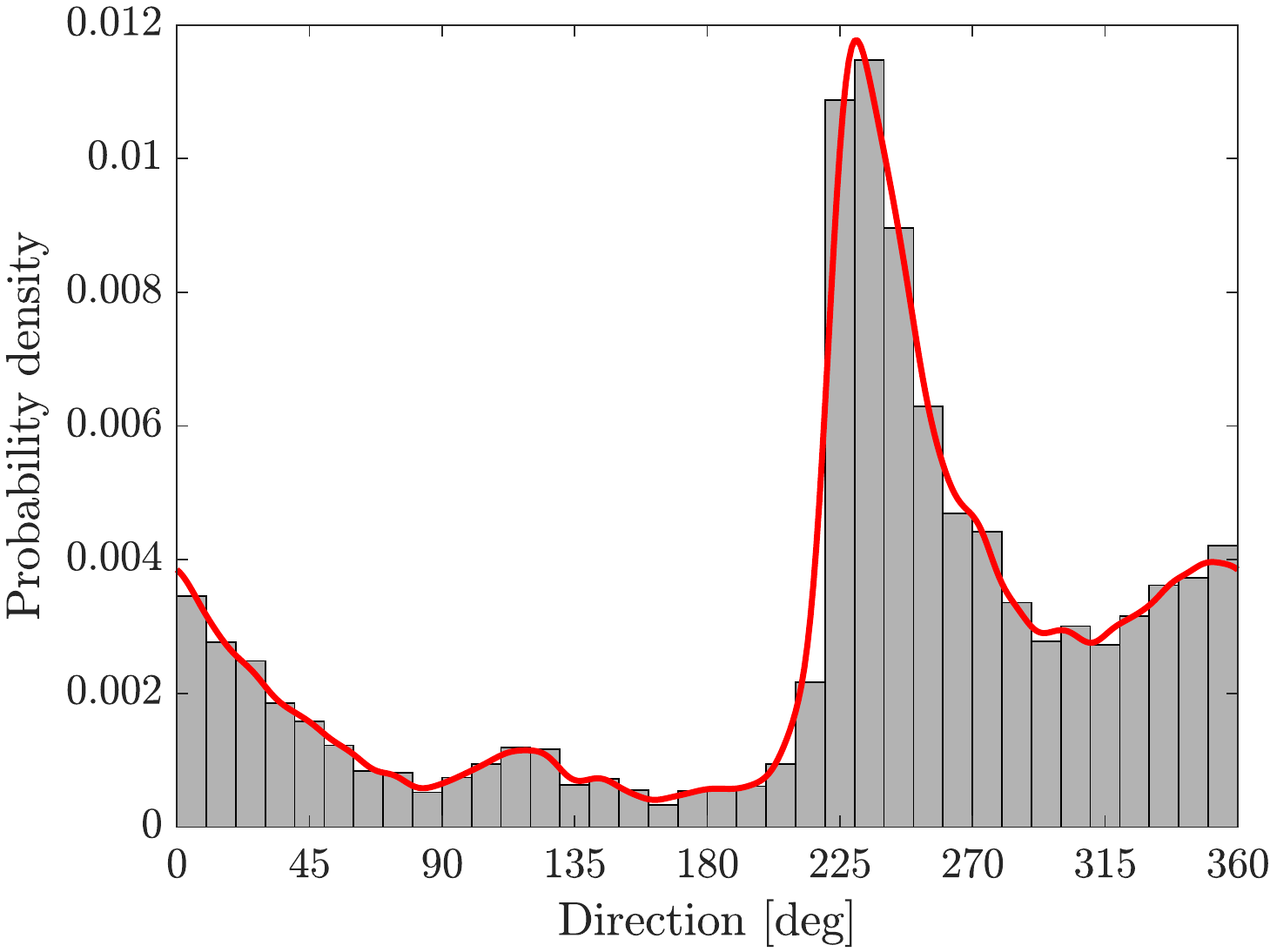}
    \captionit{Kernel density estimates (red) of covariate density for 1-D seasonal and directional analyses, with empirical histograms (grey) for comparison. Kernel bandwidth is 15 days for seasonal analysis and $5^\circ$ for directional analysis.}
    \label{fig:1cov_pdf}
\end{figure}

Kernel density estimates for the covariate density $f_{\mathbf{X}}(\mathbf{x})$ are shown in Figure~\ref{fig:1cov_pdf} for both seasonal and directional cases. After some experimentation, we chose to set the kernel density bandwidth to $w=15$ days, to achieve a reasonable seasonal model, compared to the empirical histogram. For the directional model, a bandwidth of $w=5^\circ$ was chosen to better resolve variation around $225^\circ$. 

\begin{figure}
    \centering
    \includegraphics[scale=0.5]{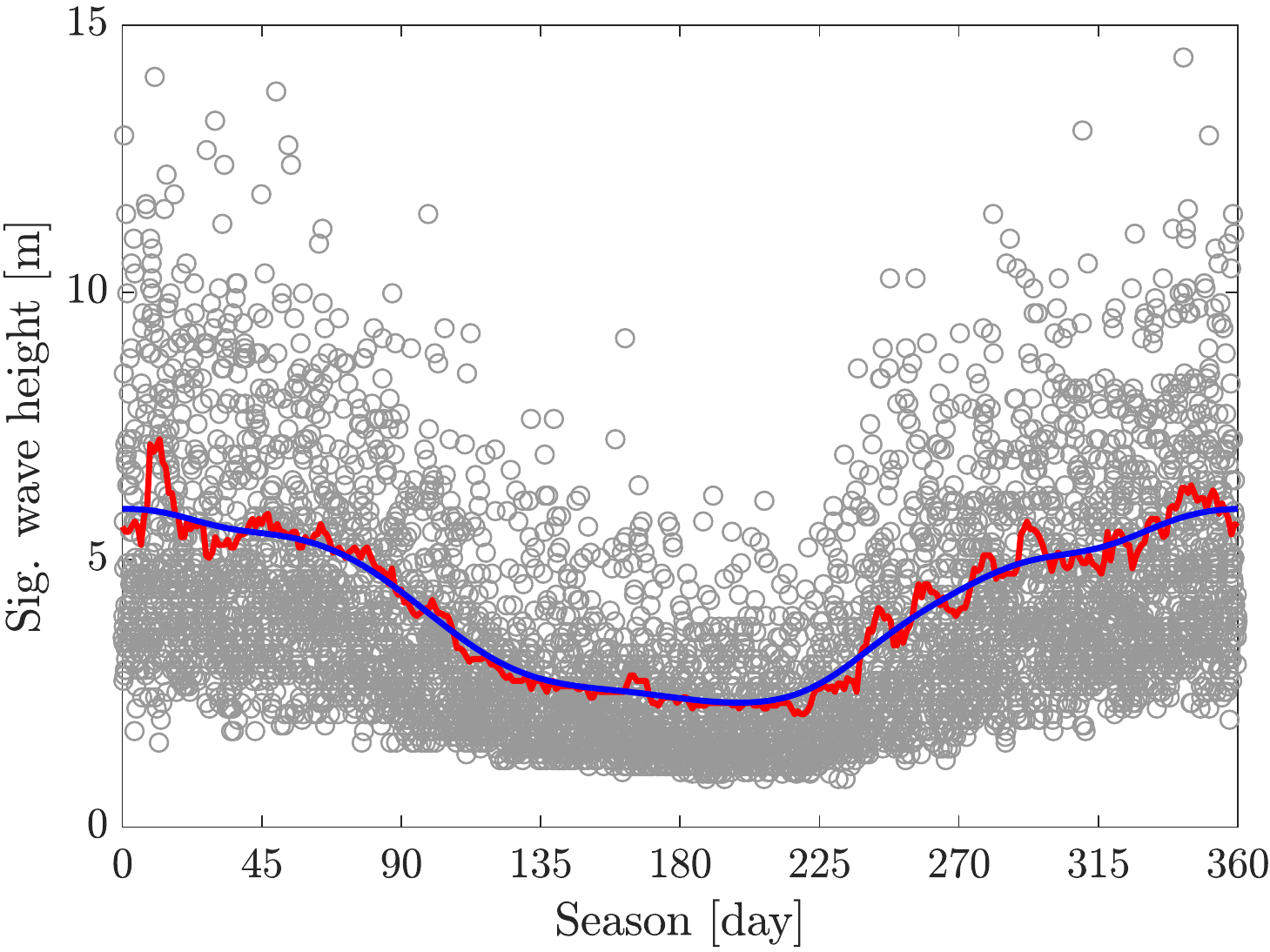} 
    \includegraphics[scale=0.5]{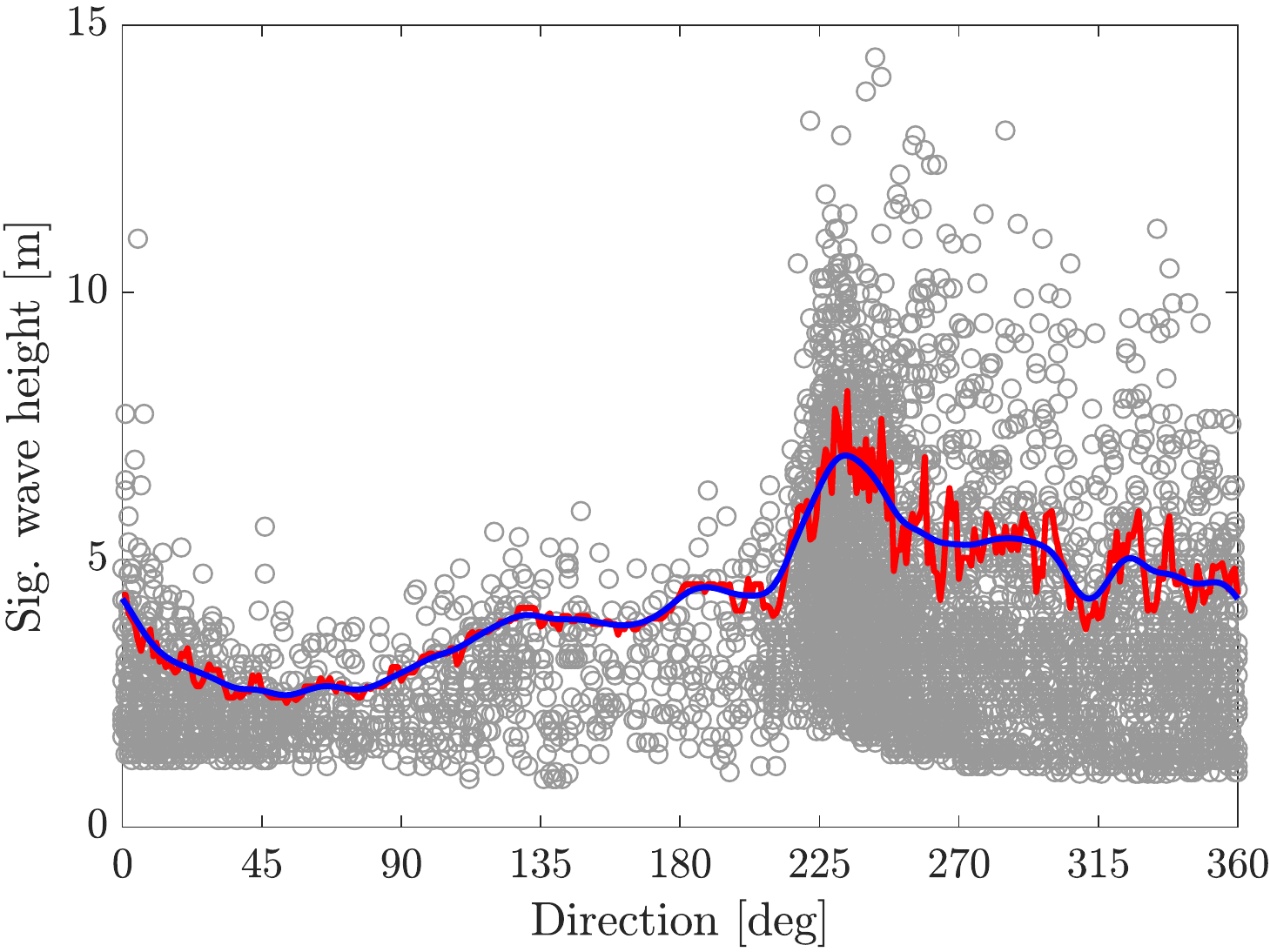}
    \captionit{Threshold estimates for 1-D seasonal and directional analyses. Preliminary local quantile estimates in red, smoothed estimates in blue. Observations are shown in circles. Threshold exceedance probability is set at $\zeta=0.3$ for seasonal analysis and $\zeta=0.2$ for directional analysis.}
    \label{fig:1cov_thresh}
\end{figure}

To estimate the non-stationary extreme value threshold $u$, the threshold exceedance probability $\zeta$ was set to 0.3 (0.2) for seasonal (or directional) analyses respectively. Inspection of model output confirmed that these choices of $\zeta$ ensure relatively large samples are retained for GP inference, without incurring too much apparent bias. Sample sizes of $N=1584$ (or 1077) threshold exceedances were retained. Estimates of the corresponding seasonal (or directional) quantiles, $u$, utilise $C=100$ (or 50) nearest observations of threshold exceedance, smoothed using a Gaussian kernel with a bandwidth of 15 days (or $5^\circ$). The larger value of $C$ and smoothing bandwidth for seasonal analysis reflects smoother variation of $H_S$ with season compared to direction. Resulting threshold estimates are shown in Figure~\ref{fig:1cov_thresh}. For the seasonal analysis, the threshold ranges from $u=2.3$\,m in summer to $5.9$\,m in winter. For the directional analysis, the threshold ranges from $u=2.5$\,m for storms from the northeast to $6.9$\,m for storms from the southwest.

\begin{figure}
    \centering
    \includegraphics[scale=0.5]{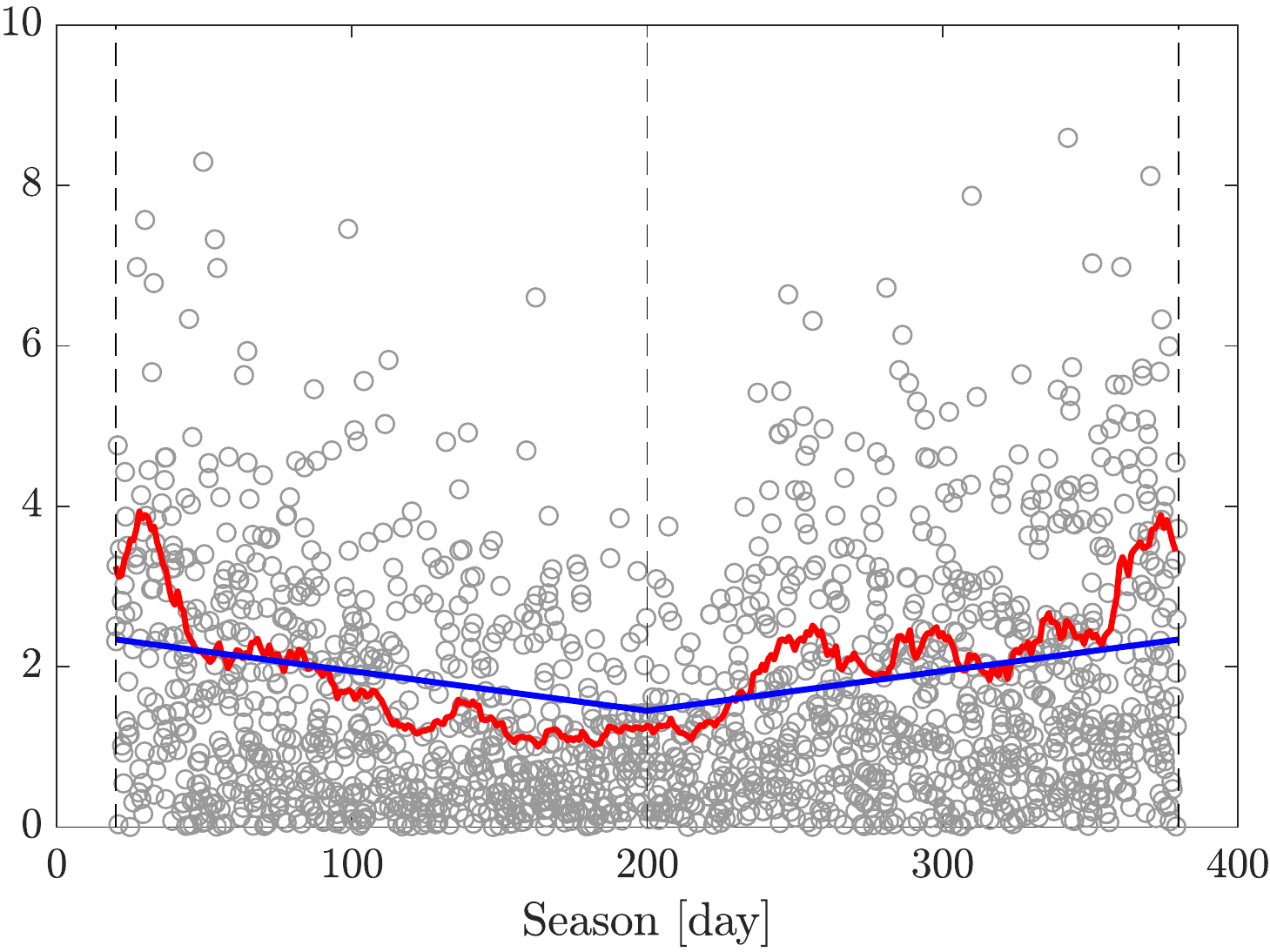} 
    \includegraphics[scale=0.5]{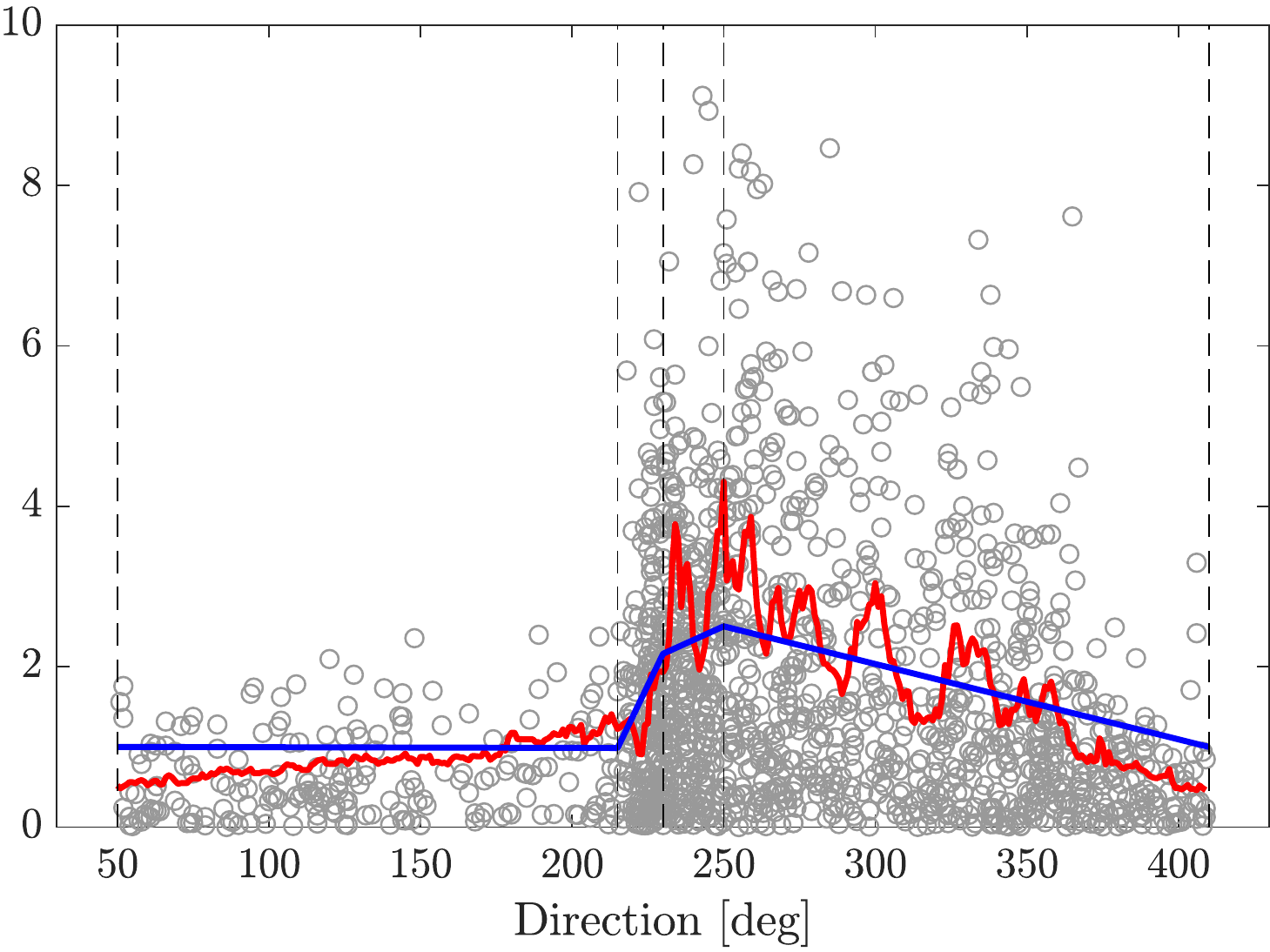}\\
    \includegraphics[scale=0.5]{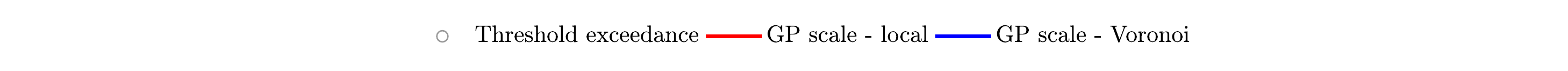}
    \captionit{Threshold exceedances together with initial local estimates of GP scale parameter based on nearest observations. Starting values for the PPL estimates of scale based on the Voronoi partition are also shown. The locations of the nodes are shown as vertical dashed lines.}
    \label{fig:1cov_voronoi}
\end{figure}

Seasonal and directional models with $K=2$, 3, 4 and 6 nodes are estimated, corresponding to each of Cases A, B and C above. For the seasonal analysis, nodes are spaced uniformly on covariate domain $\mathcal{D}$, with the first node location at $n_1=20$ days, corresponding to the peak value of the initial local estimate of GP scale $\sigma$. For the directional case we place nodes manually, starting with 2 nodes located at the minimum and maximum of the initial local estimates of $\sigma$. For increasing $K$, additional nodes are added in turn, without changing the locations of existing nodes. Examples of the node placement for the $K=2$ seasonal model and $K=4$ directional model are shown in Figure~\ref{fig:1cov_voronoi}, together with the local estimates (red) of $\sigma$, and corresponding initial values (blue) for the PPL model from a Voronoi partition. The blue curve appears to provide a reasonable representation of the variation in scale apparent in the local red estimates.

\begin{figure}
    \centering
    \includegraphics[scale=0.5]{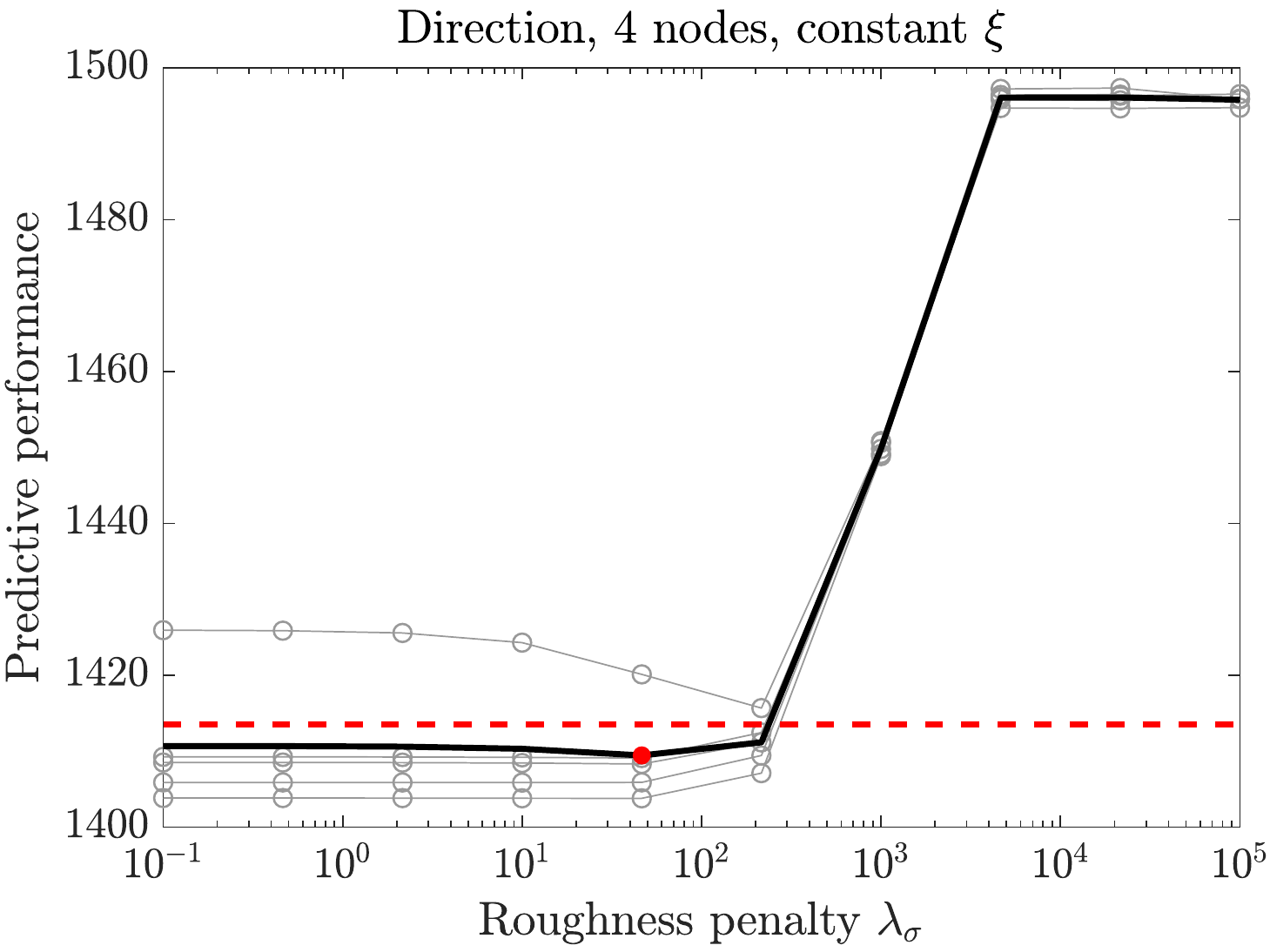}
    \includegraphics[scale=0.5]{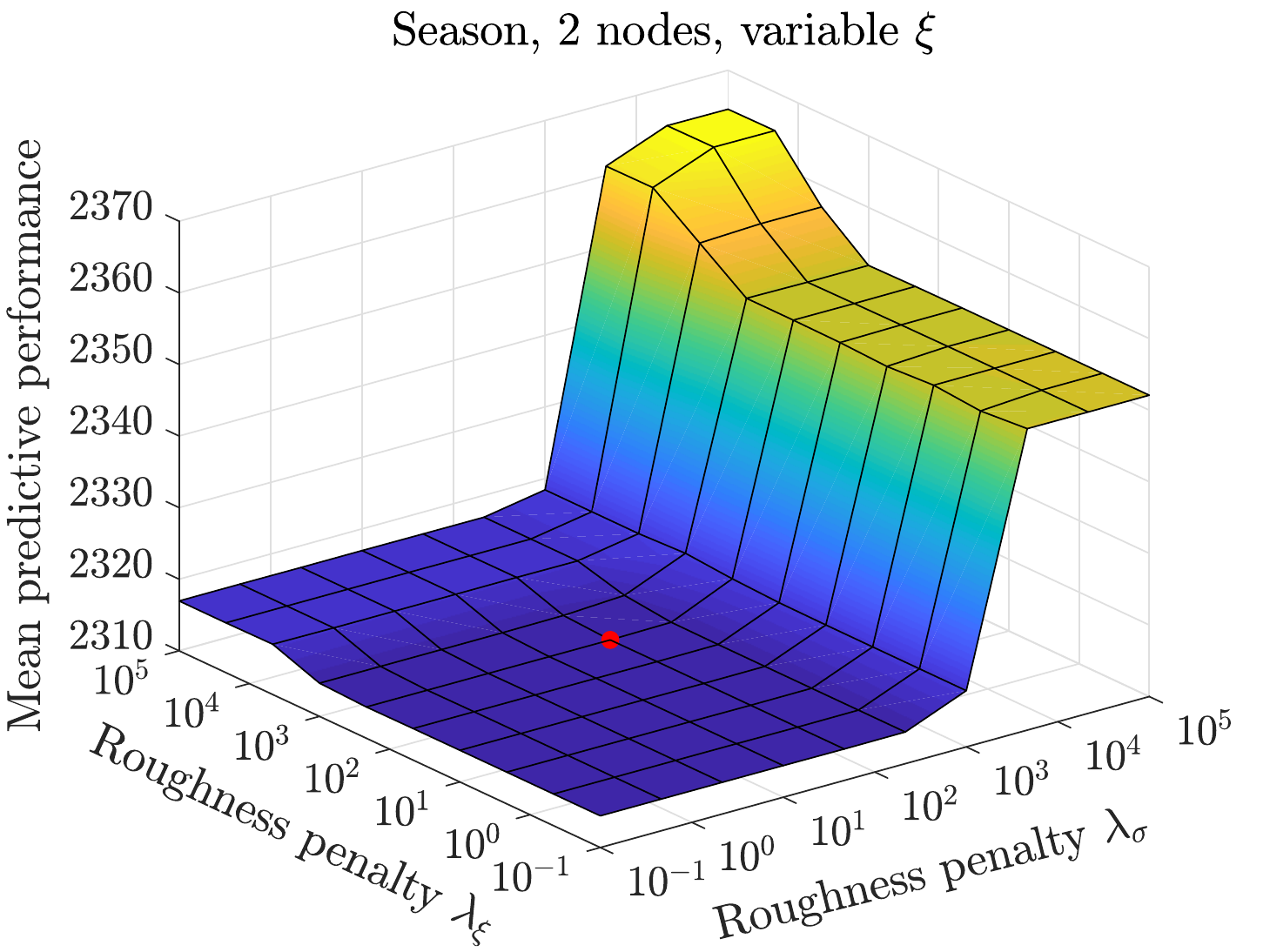}
    \captionit{Optimal roughness coefficients for 1-D models. Left: Predictive performance (negative log-likelihood) against roughness coefficient $\lambda_{\sigma}$, for directional Case A with $K=4$. Grey lines indicate results for $R=5$ replicate runs, and the solid black line is the mean over replicates. The dashed red line shows the value of the optimal predictive performance $\bar{\mathcal{P}}(\boldsymbol{\lambda}^*)$ recorded. Right: Mean predictive performance over $R=5$ replicates for seasonal Case B with $K=2$, with roughness coefficients $\lambda_{\xi}$, $\lambda_{\sigma}$. In both panels, the red dot shows the location of penalty vector $\boldsymbol{\lambda}^\circ$ yielding optimal predictive performance.}
    \label{fig:1cov_crossval}
\end{figure}

The left panel of Figure~\ref{fig:1cov_crossval} illustrates the choice of $\lambda_\sigma^\circ$ for directional Case A with $K=4$ nodes. The minimum mean negative log predictive likelihood $\bar{\mathcal{P}}(\lambda_{\sigma}^\circ)$ occurs at $\lambda_{\sigma}=10^{1.67}$. However, values of $\lambda_{\sigma}\leq 10^{2.33}$ are all within the range of acceptable predictive performance, as quantified by the jackknife uncertainty in $\bar{\mathcal{P}}(\lambda_{\sigma}^\circ)$. Hence we set the optimal penalty as $\lambda_{\sigma}^* = 10^{2.33}$ for this case. The utility of replicating the cross-validation procedure $R=5$ times is evident: for four of five repeats, the predictive performance is relatively constant up to the optimal penalty, whereas on one repeat, the performance deteriorates for $\lambda_{\sigma}<\lambda_{\sigma}^*$. The sharp increase in $\bar{\mathcal{P}}(\lambda_{\sigma})$ around $\lambda_{\sigma}=10^3$ indicates that the corresponding models are too stiff to capture the variation of the scale parameter. For $\lambda_{\sigma}\geq 10^{3.67}$, predictive performance asymptotes as the estimated GP scale $\sigma$ is effectively constant on $\mathcal{D}$.

The corresponding plot for seasonal Case B with $K=2$ nodes is shown in the right panel. Variation of mean predictive performance $\bar{\mathcal{P}}$ with $\lambda_{\sigma}$ is similar to that in the left panel. There is less variation in $\bar{\mathcal{P}}$ with $\lambda_{\xi}$, indicating that allowing $\xi$ to vary on $\mathcal{D}$ provides only small improvement in model performance.

For case A, values of optimal predictive performance $\bar{\mathcal{P}}(\boldsymbol{\lambda}^*)$, associated uncertainties and optimal values for roughness coefficients $\boldsymbol{\lambda}^*$ for the 1-D seasonal models with $K=2, 3, 4$ and $6$ are given in Tables \ref{tab:CV_season}. Values of predictive performance scores are directly comparable between seasonal models, since all are based on the same sample.
\begin{table}
\centering
	\begin{tabular}{l | l | c c c c}
	& & \multicolumn{4}{c}{Number of nodes, $K$}\\
	Case & Variable & 2 & 3 & 4 & 6\\
	\hline
	\multirow{2}{*}{A} & Optimal performance & 2316 (0.7) & 2321 (1.3) & 2314 (2.1) & 2324 (0.6)\\
     & $\log_{10}\left(\lambda_{\sigma}^*\right)$ & 2.3 & 2.3 & 2.3 & 3\\
     \hline
    \multirow{3}{*}{B} & Optimal performance & 2314 (2.0) & 2316 (3.3) & 2311 (1.2) & 2308 (2.5)\\
    & $\log_{10}\left(\lambda_{\sigma}^*\right)$ & 2.3 & 2.3 & 1.7 & 2.3\\
    & $\log_{10}\left(\lambda_{\xi}^*\right)$ & 3 & 3 & 3 & 2.3
	\end{tabular}
	\captionit{Optimal predictive performance, $\bar{\mathcal{P}}(\boldsymbol{\lambda}^*)$, and optimal penalties, $\boldsymbol{\lambda}^*$, for 1-D seasonal model with $K=2$, 3, 4 and 6 nodes. Numbers in brackets are approximate uncertainties on performance from jackknife analysis.}
	\label{tab:CV_season}
\end{table}

Although there is some variability in $\bar{\mathcal{P}}(\boldsymbol{\lambda}^*)$, given its uncertainty we can conclude that model performance is generally independent of $K$. Performance at $K=6$ is somewhat worse, probably because that analysis suggests that a larger value of $\lambda_{\sigma}^*=10^3$ is appropriate. Further investigation revealed that for smaller values of roughness coefficient, optimal parameter estimates from a particular cross-validation on sample $\mathcal{S} \setminus \mathcal{S}_g$ (i.e. omitting subset $\mathcal{S}_g$ for some $g=1,2,...,G$) yielded a predicted likelihood of zero on prediction set $\mathcal{S}_g$, or infinite predictive negative log-likelihood; this is a common problem with prediction from responses defined on bounded domains (e.g. \citealt{NrtEA15}). Thus, flexible models tend to be rejected, and a large roughness coefficient is required for plausible predictive performance; we surmise that the 6-node model with $\lambda_{\sigma}^*<10^3$ induces some over-fitting.

\begin{figure}
    \centering
    \includegraphics[scale=0.5]{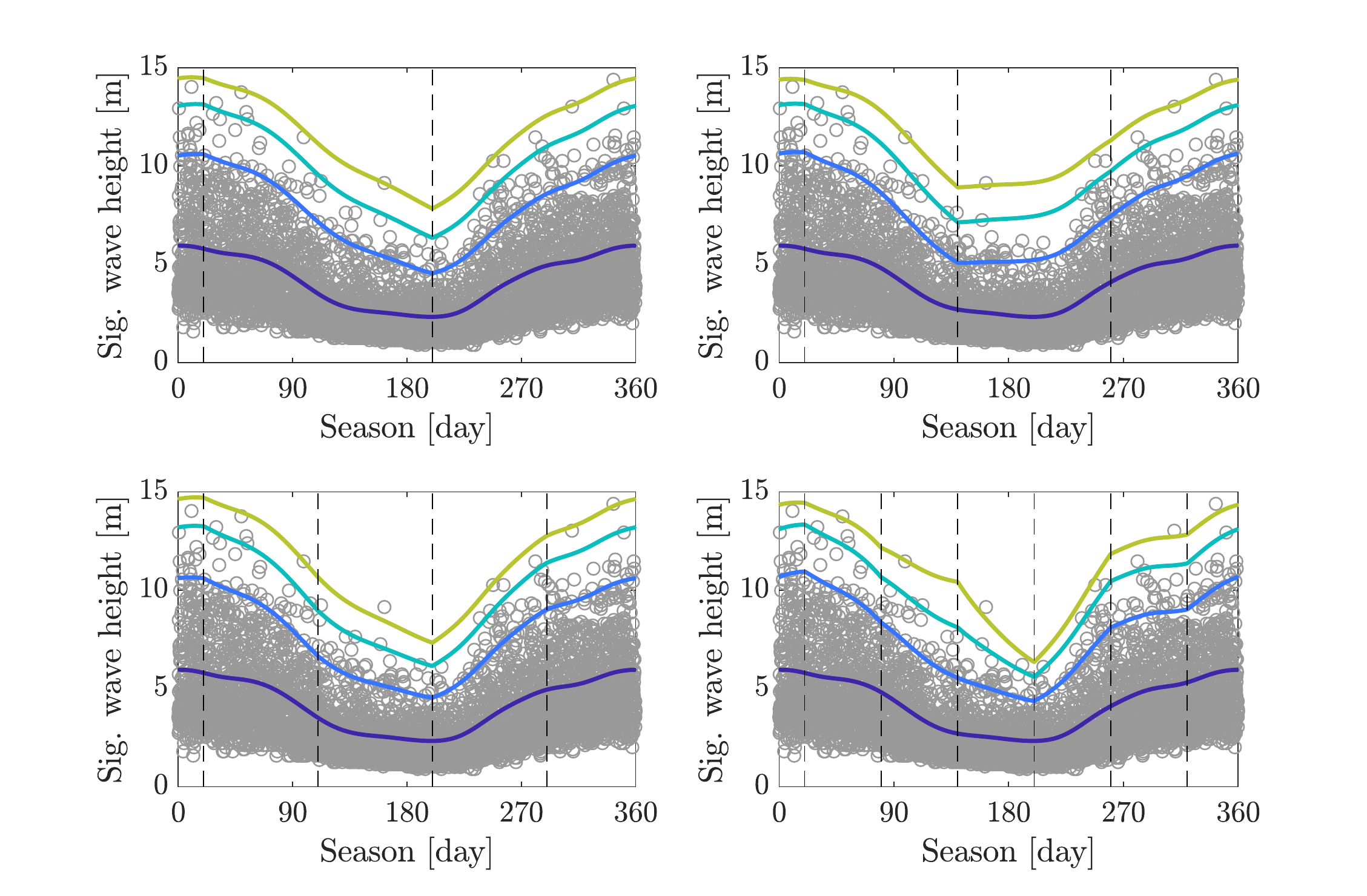}
    \captionit{Conditional quantiles for 1-D seasonal Case B model with $K=2, 3, 4$ and 6. Coloured lines show quantiles at non-exceedance probabilities of 0.7 (corresponding to threshold, $u$), 0.9, 0.99 and 0.999. Circles are original observations. Vertical lines show node locations.}
    \label{fig:1cov_season_quant}
\end{figure}

There is a slight improvement in performance, relative to the uncertainty associated with the predictive performance estimates, for Case B over Case A. We note that, generally, a larger roughness coefficient is selected for GP shape $\xi$ than for scale $\sigma$. Conditional quantiles corresponding to different non-exceedance probabilities, estimated under the fitted model for different $K$, are shown in Figure~\ref{fig:1cov_season_quant}. Unsurprisingly, given the similar predictive performances listed in Table~\ref{tab:CV_season}, quantile levels are similar for all choices of $K$, with $K=6$ suggesting somewhat lower extreme quantile levels in the interval [140, 200] days.

\begin{figure}
    \centering
    \includegraphics[scale=0.5]{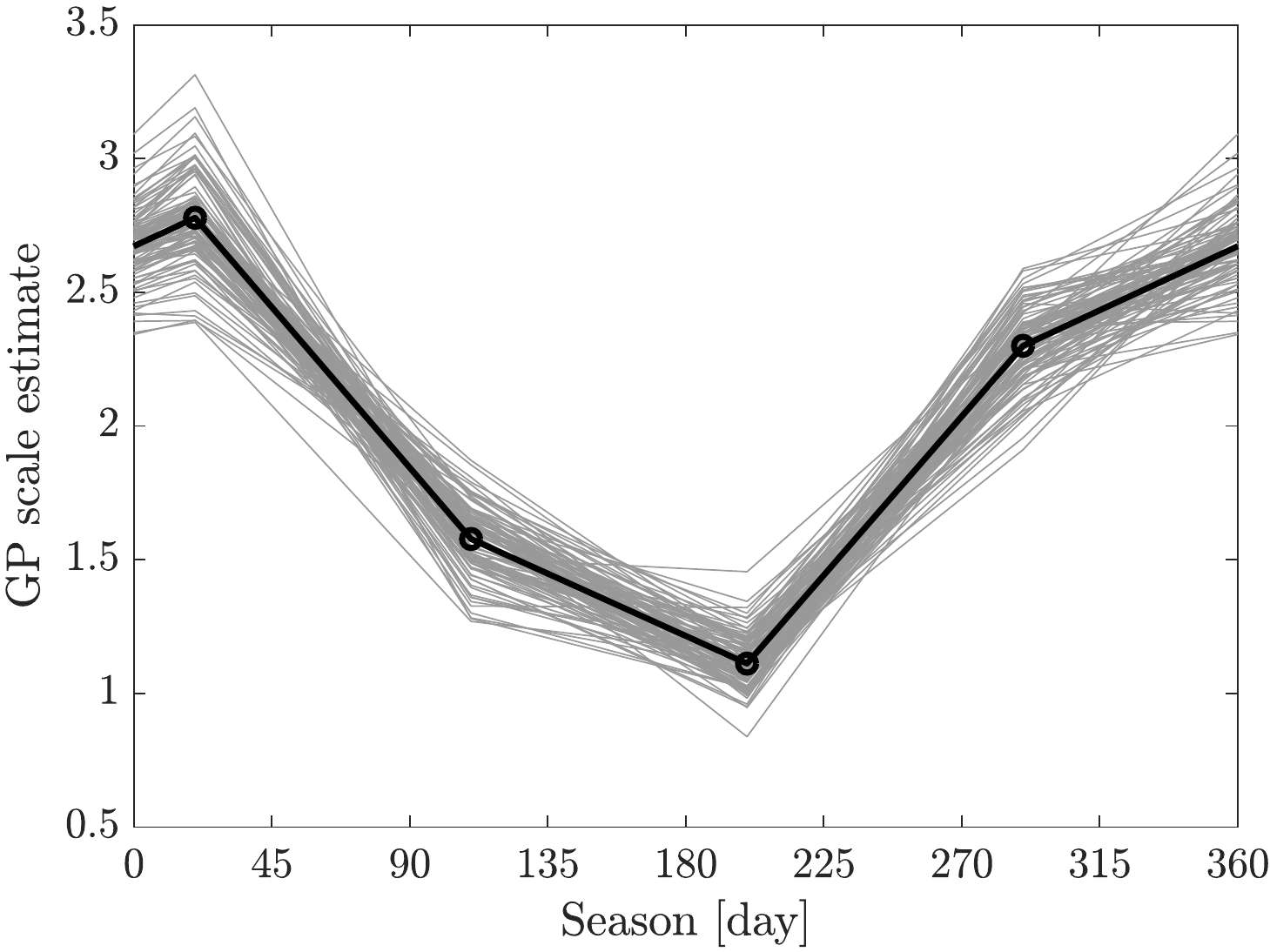}
    \includegraphics[scale=0.5]{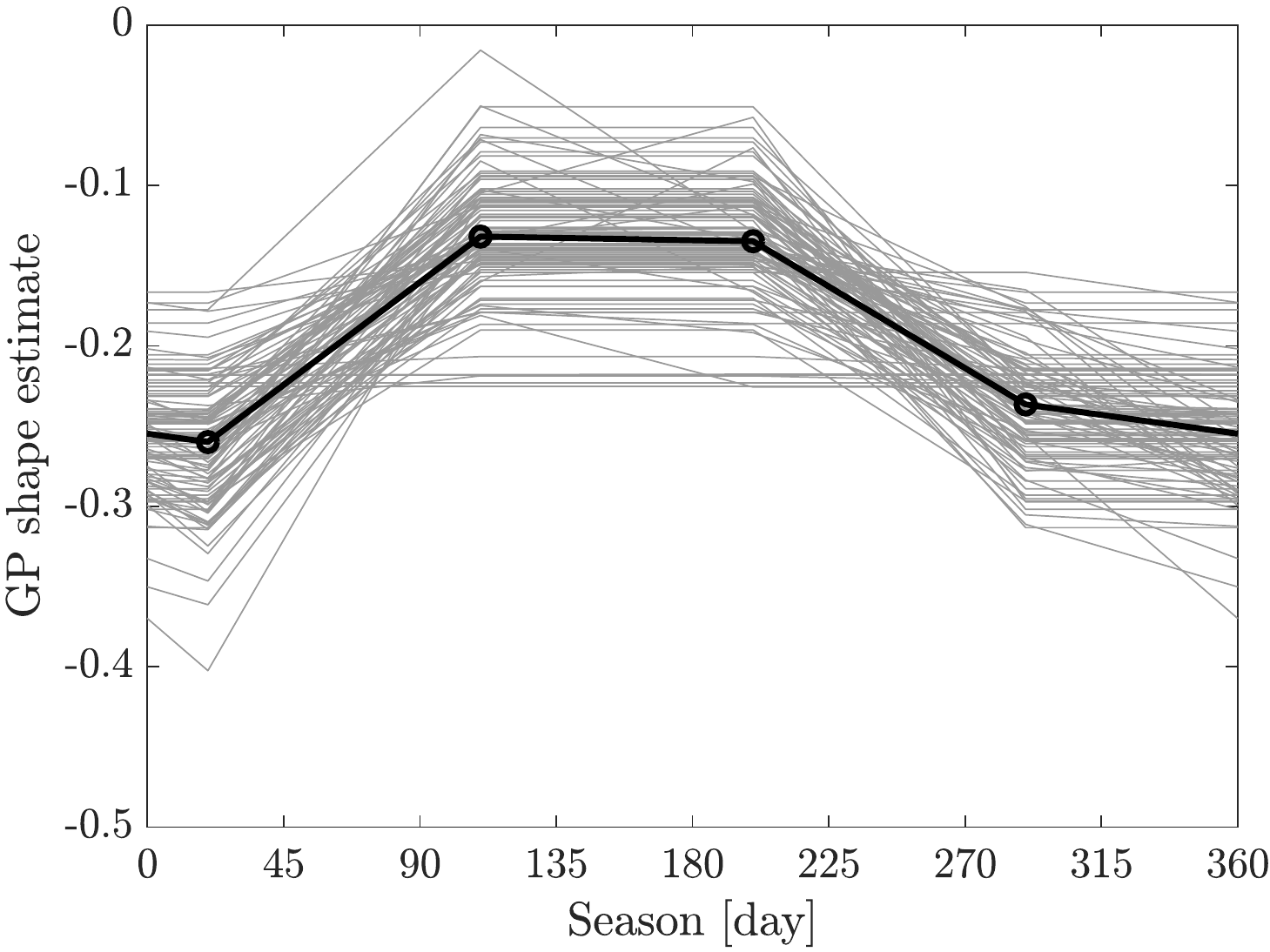}
    \captionit{Estimates of GP scale and shape parameters for 1-D seasonal Case B model with $K=4$. Grey lines show estimates from each of 100 bootstrap resamples of the original sample. Bold lines show bootstrap means.}
    \label{fig:1cov_season_param}
\end{figure}

Estimates of GP scale and shape for Case B with $K=4$ are shown in Figure~\ref{fig:1cov_season_param}, for each of 100 bootstrap resamples of the original sample $\mathcal{S}$. Parameter estimates for individual bootstrap resamples are shown in grey, and their mean in black. Seasonal variation is clear, with larger scale in winter and somewhat more positive shape in the summer. Uncertainty in the scale parameter estimate is much lower than the magnitude of the seasonal variation. The corresponding uncertainty in the shape parameter estimate is higher relative to the seasonal variability; for some bootstrap resamples, the estimated shape parameter was approximately constant with season (visible as the straight horizontal lines). 

\begin{figure}
    \centering
    \includegraphics[scale=0.5, bb=80 10 970 525, clip=true]{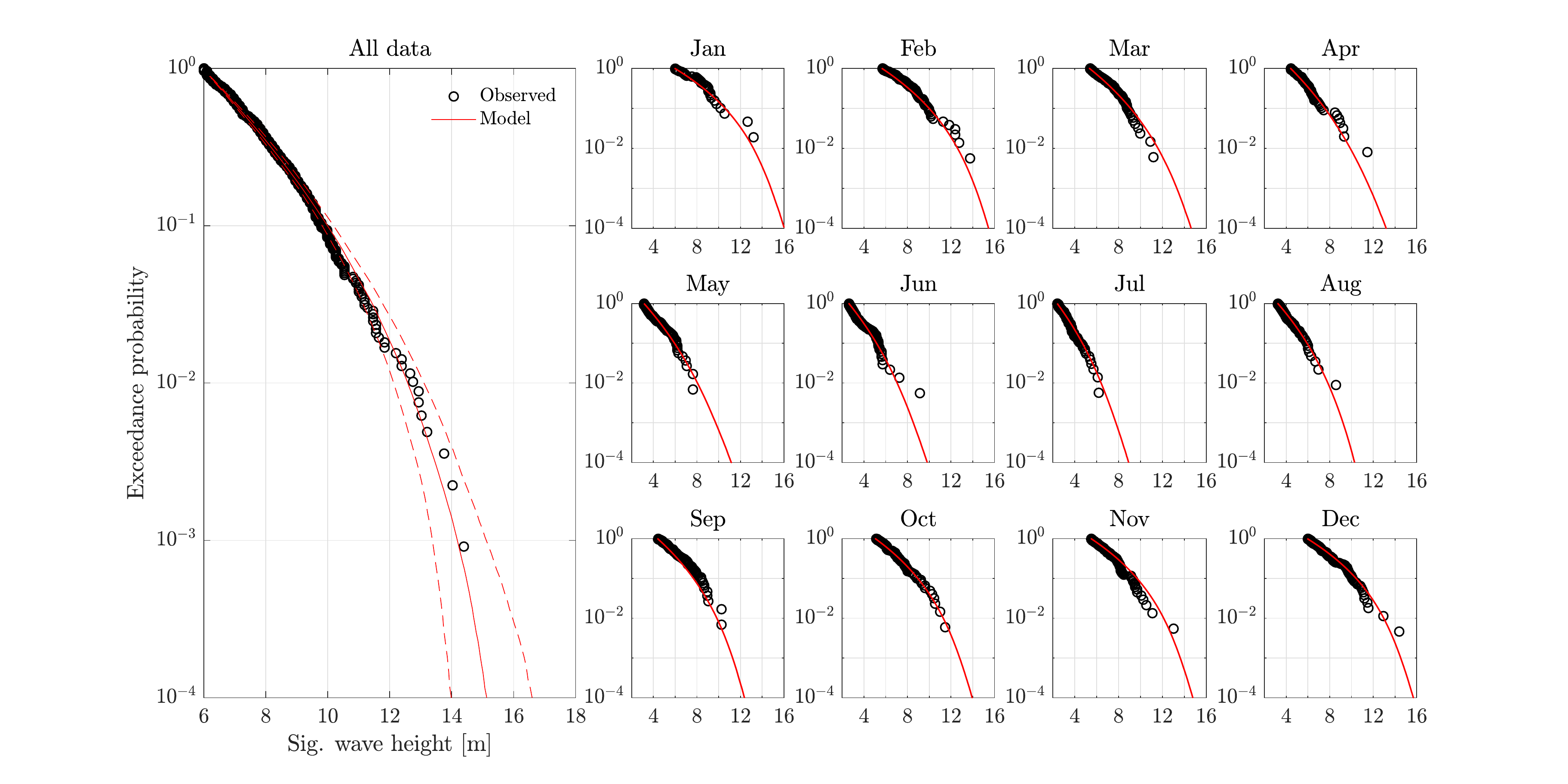}
    \captionit{Annual and monthly exceedance probabilities for the 1-D seasonal Case B model with $K=4$. Empirical estimates in black, and model-based estimates in red. Dashed lines on annual exceedance probability plot represent 95\% uncertainty bands based on 100 bootstrap trials.}
    \label{fig:1cov_season_exceed}
\end{figure}

Comparisons of the tail probabilities $\Pr(Y > y)$ estimated directly from the sample, and from simulation under the fitted 1-D seasonal Case B model with $K=4$ are shown in Figure~\ref{fig:1cov_season_exceed}. Corresponding ``monthly'' tails are also compared. The number of the simulated points is $1000 \times N$, (where $N$ is the original sample size). This reduces sampling uncertainty in the empirical distribution function for the simulated data to a reasonable level at the exceedance probability associated with the largest observation (see \citealt{Mackay2021sampling} for details). For the left-hand all-year plot, a 95\% bootstrap uncertainty band (based on 100 bootstrap resamples) is also shown; for clarity, uncertainty bands on monthly plots are suppressed). There is good agreement between observations and the fitted model. 

\begin{figure}
    \centering
    \includegraphics[scale=0.5]{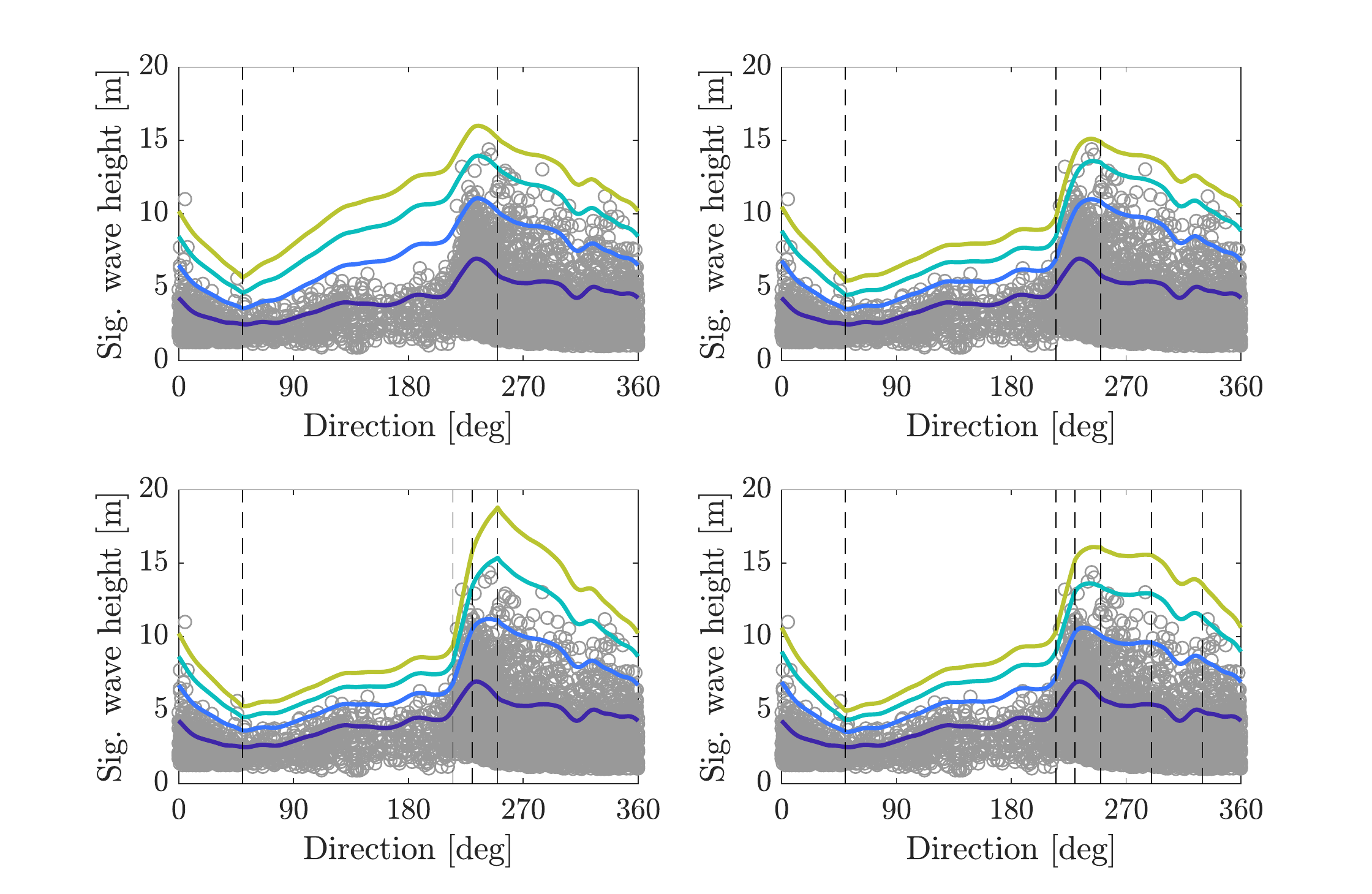}
    \captionit{Conditional quantiles for 1-D directional Case B model with $K=2, 3, 4$ and 6. Coloured lines show quantiles at non-exceedance probabilities of 0.8 (corresponding to threshold, $u$), 0.9, 0.99 and 0.999. Circles are original observations. Vertical lines show node locations. Note that for $K=4,6$, optimal models correspond to constant GP shape.}
    \label{fig:1cov_dir_quant}
\end{figure}

\begin{figure}
    \centering
    \includegraphics[scale=0.5]{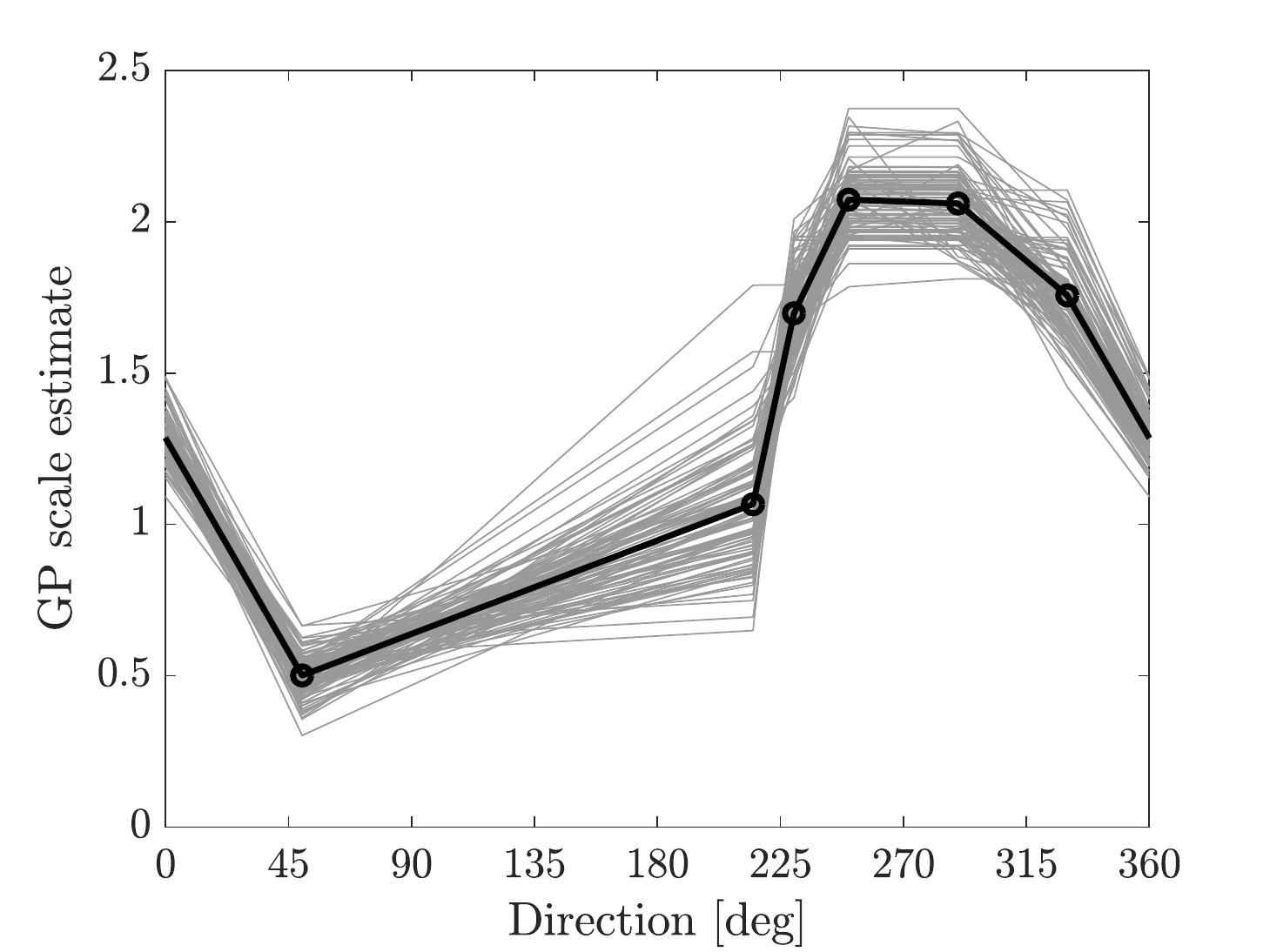}
    \includegraphics[scale=0.5]{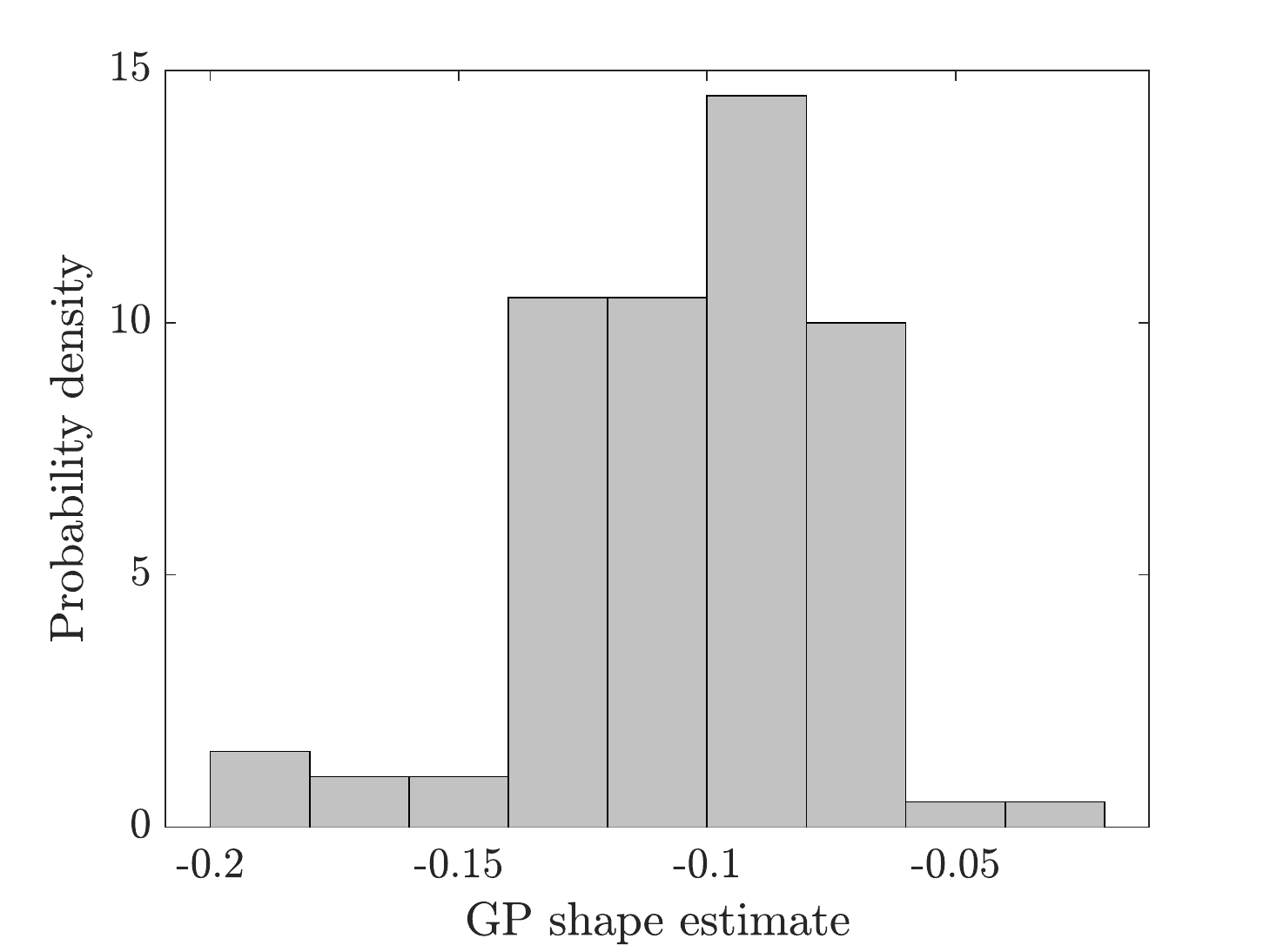}
    \captionit{Estimates of GP scale and shape parameters for 1-D directional Case A model with $k=6$. In the left panel, grey lines show estimates from each of 100 bootstrap resamples of the original sample, and bold lines show bootstrap means. The right panel gives a histogram of the corresponding shape estimates.}
    \label{fig:1cov_dir_param}
\end{figure}

Corresponding results for 1-D directional models for Cases A and B are summarised in Table~\ref{tab:CV_dir}.
\begin{table}
\centering
	\begin{tabular}{l | l | c c c c}
	& & \multicolumn{4}{c}{Number of nodes, $K$}\\
	Case & Variable & 2 & 3 & 4 & 6\\
	\hline
	\multirow{2}{*}{A} & Optimal performance & 1438 (0.7) & 1410 (2.6) & 1409 (4.1) & 1409 (2.2)\\
    & $\log_{10}\left(\lambda_{\sigma}^*\right)$ & 2.3 & 2.3 & 2.3 & 1.7\\
     \hline
    \multirow{3}{*}{B} & Optimal performance & 1432 (0.7) & 1403 (1.5) & 1406 (2.0) & 1413 (5.2)\\
    & $\log_{10}\left(\lambda_{\sigma}^*\right)$ & 2.3 & 1.7 & 1.7 & 2.3\\
    & $\log_{10}\left(\lambda_{\xi}^*\right)$ & 2.3 & 1.7 & 5 & 5
	\end{tabular}
	\captionit{Optimal predictive performance, $\bar{\mathcal{P}}(\boldsymbol{\lambda}^*)$, and optimal penalties, $\boldsymbol{\lambda}^*$, for 1-D directional model with $K=2$, 3, 4 and 6 nodes. Numbers in brackets are approximate uncertainties on performance from jackknife analysis.}
	\label{tab:CV_dir}
\end{table}
There is little difference in performance for 1-D directional models for Cases A and B, with $K>2$. However, $K=2$ suggests poorer performance, indicating that at least three nodes are required to capture directional variability. Allowing shape to vary gives a small improvement. However, for Case B with $K=4,6$ the optimal roughness coefficients correspond to effectively constant shape on $\mathcal{D}$, again resulting from occurrences of infinite negative log predictive likelihoods for smaller choices of roughness coefficient. Conditional quantiles corresponding to different non-exceedance probabilities, estimated under the fitted Case B model for different $K$, are shown in Figure~\ref{fig:1cov_dir_quant}; $K=2$ does not adequately capture behaviour in the land shadow of Norway at angles in the interval $[50,200]$; for larger values of $K$, quantile levels are generally comparable. For comparison with Figure~\ref{fig:1cov_season_param}, estimates of GP scale and shape for Case A with $K=6$ are shown in Figure~\ref{fig:1cov_dir_param}, for each of 100 bootstrap resamples of the original sample, $\mathcal{S}$.

Exceedance probability plots were generated for all 1-D models.  The plots for the directional models showed comparable levels of agreement to that illustrated in Figure~\ref{fig:1cov_season_exceed}, so are not shown here. Note that the predictive performance of the directional and seasonal models cannot be compared, as the set of threshold exceedances is different in each case, due to the different choices of directional and seasonal non-stationary thresholds, with different exceedance probabilities, $\zeta$. 

\section{Application with two-dimensional covariate} \label{sec:App:2D} %
Here we discuss the estimation of two-dimensional directional-seasonal models, again following the procedure outlined in Section~\ref{sec:MdlFrm:Jnt}. Note that the definition of the Cases A, B, C archetypes is given in Section~\ref{sec:App:CmpCmp}.

\begin{figure}
    \centering
    \includegraphics[scale=0.5]{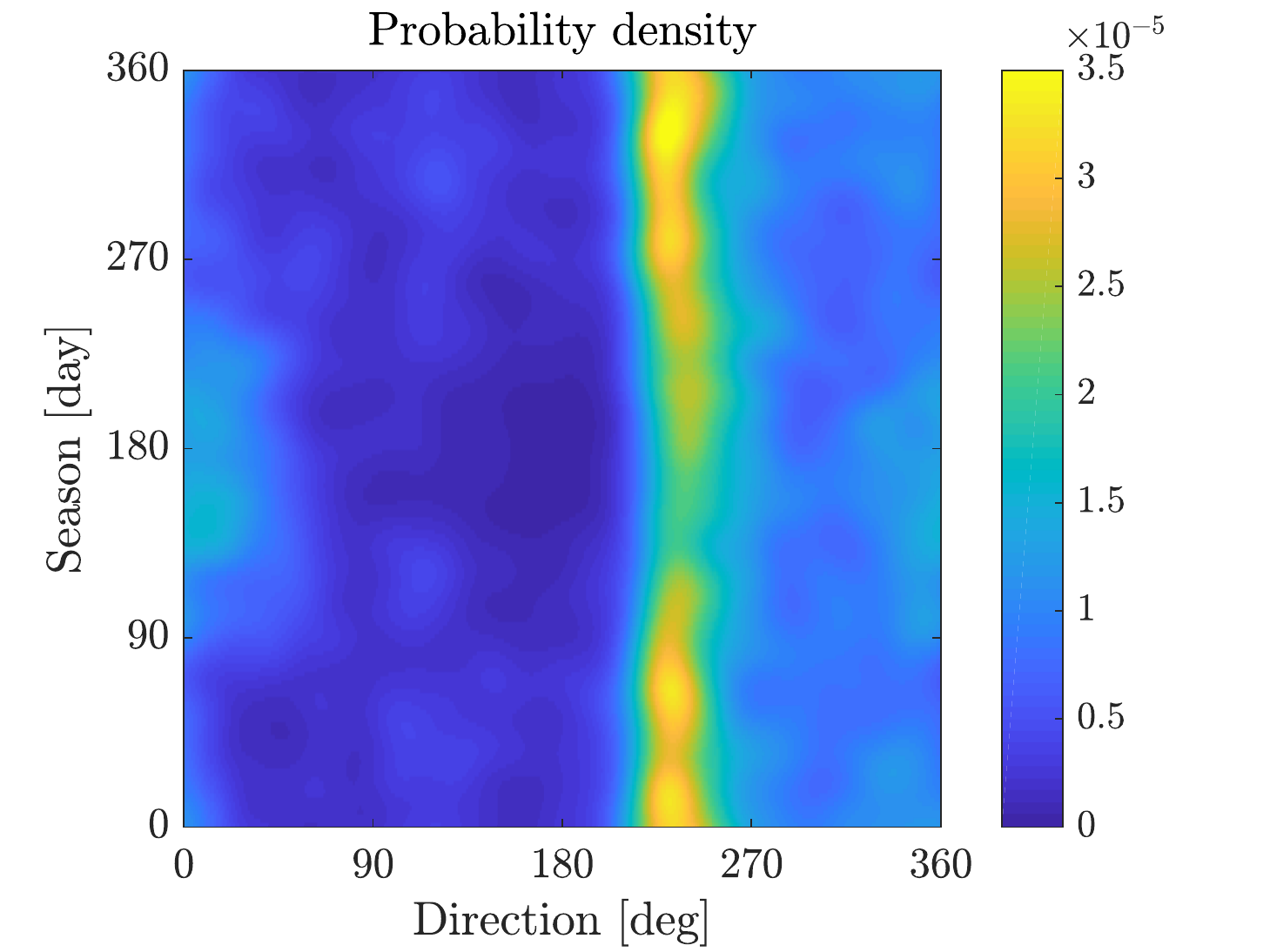}
    \includegraphics[scale=0.5]{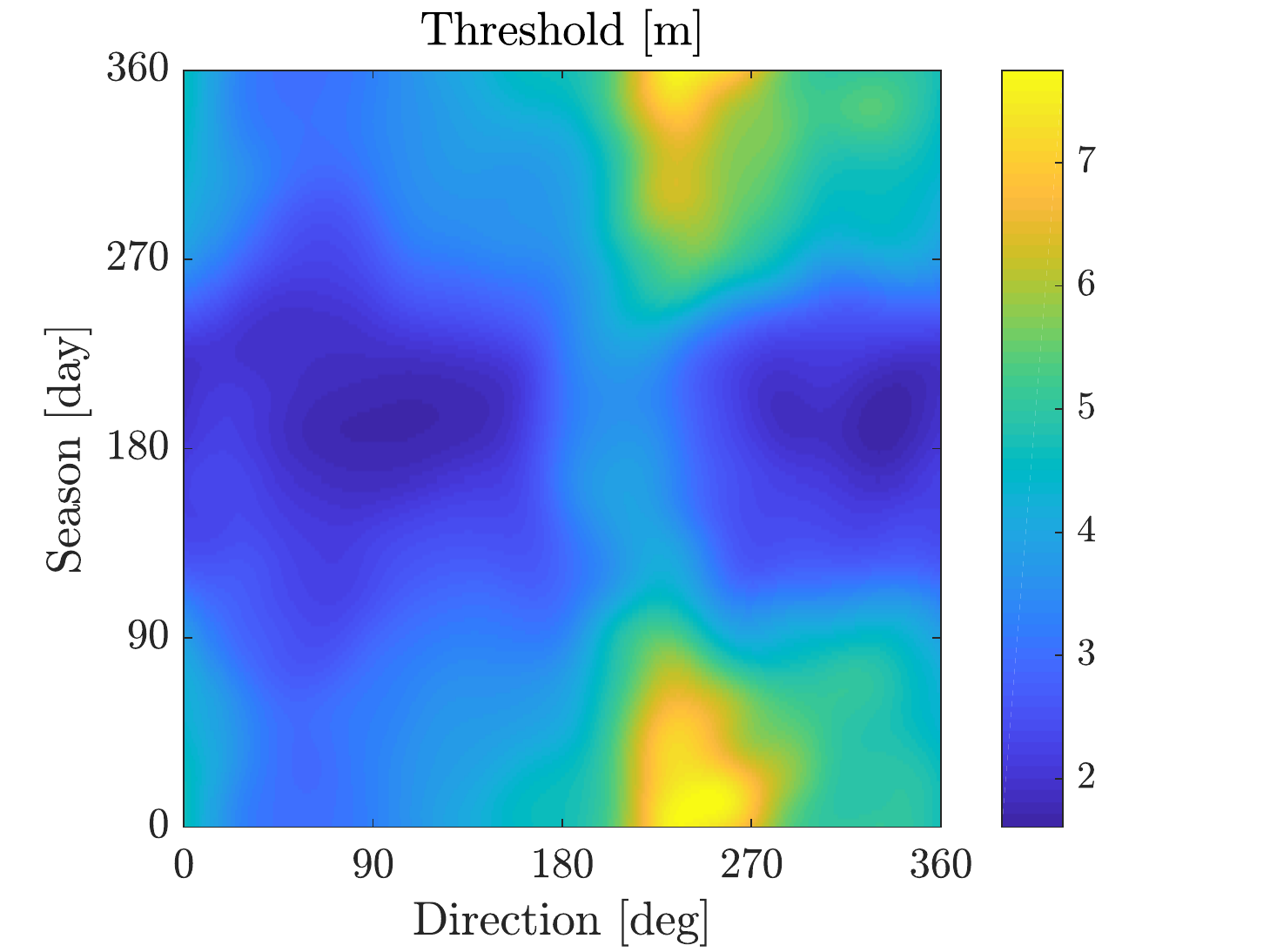} 
    \captionit{Estimates of covariate density (left) and GP threshold parameter (right) for 2-D directional-seasonal analysis. Kernel bandwidth $w=10$ degrees (directionally) and days (seasonally) for both kernel density estimate (left) and threshold smoothing (right). Initial local threshold estimates based on 50 nearest observations.}
    \label{fig:2cov_pdf_thresh}
\end{figure}

Kernel density estimates for the covariate density $f_{\mathbf{X}}(\mathbf{x})$ are shown in the left panel of Figure~\ref{fig:2cov_pdf_thresh}, using a common kernel density bandwidth to $w=10^\circ$ directionally and $10$ days seasonally, ensuring that directional details can be resolved.
The estimated threshold corresponding to a local exceedance probability $\zeta=0.3$ is given in the right panel of the Figure~\ref{fig:2cov_pdf_thresh}, using the $C=50$ nearest  neighbours, smoothed with a Gaussian kernel with bandwidth $10$ degrees (directionally) and 10 days (seasonally).

\begin{figure}
    \centering
    \includegraphics[scale=0.5]{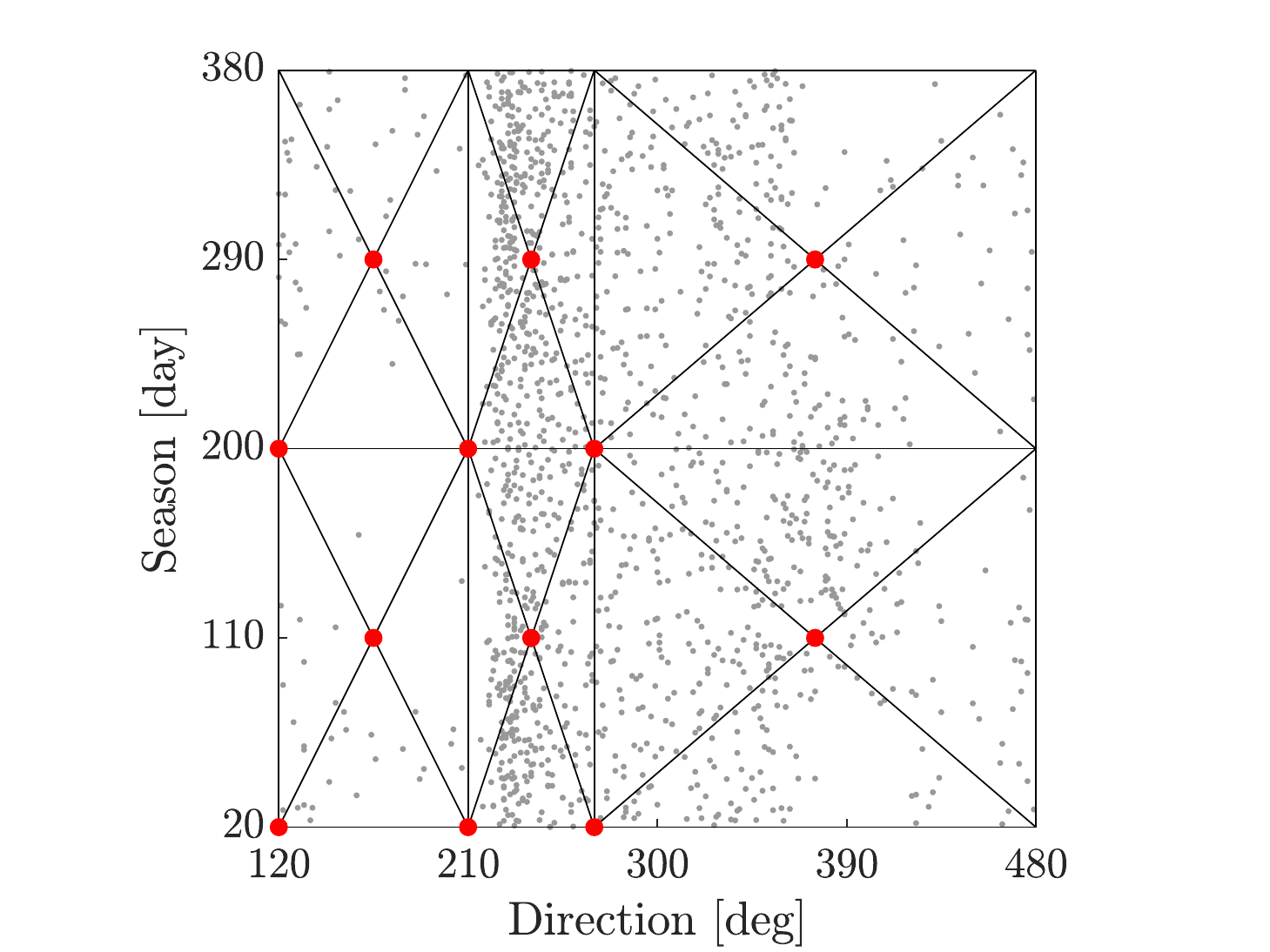}
    \includegraphics[scale=0.5]{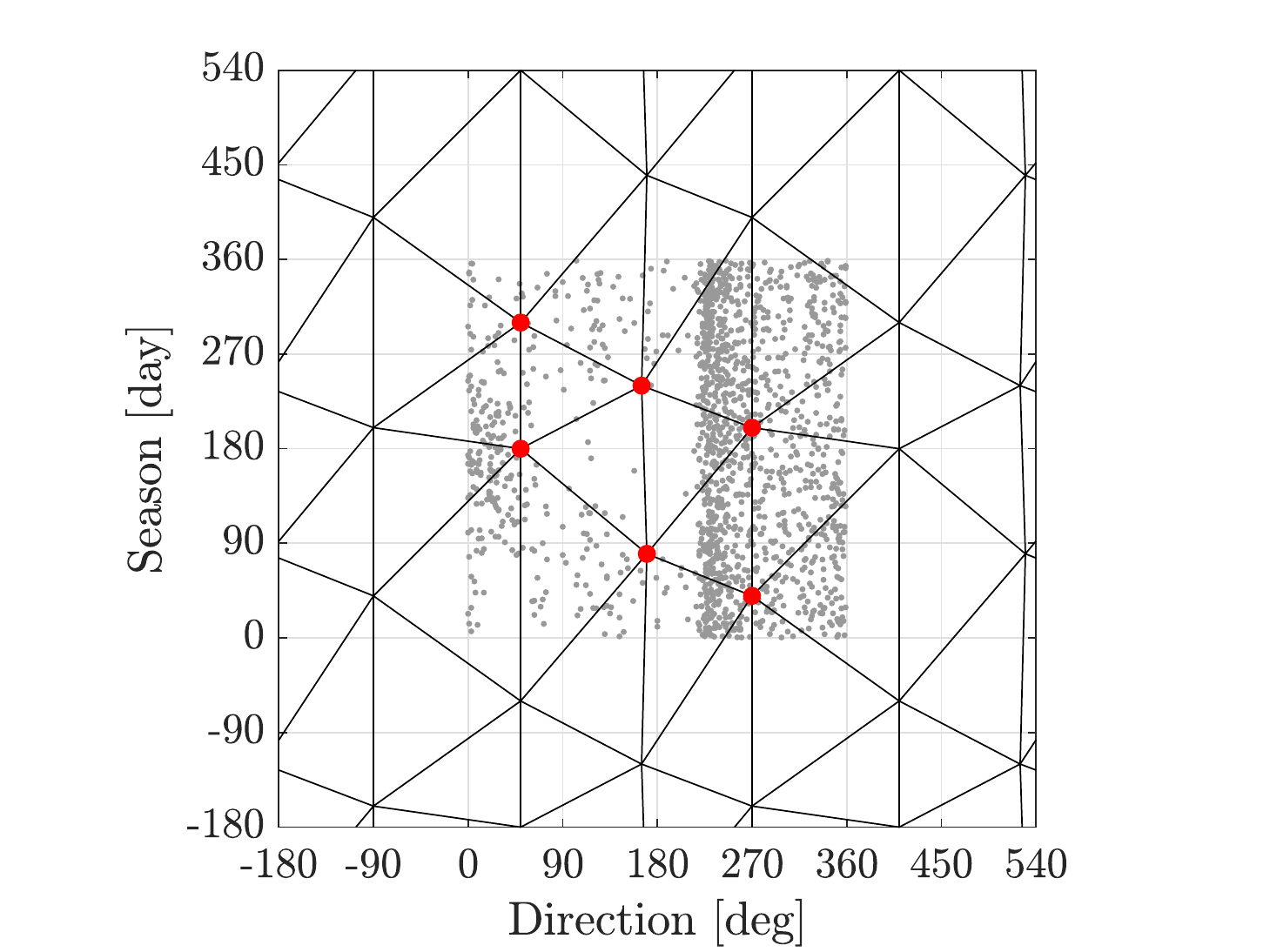} 
    \captionit{Examples of regular and irregular 2-D grids (left and right respectively) on extended covariate domain. Unique nodes are shown as red dots; all other nodes are repeats, included for ease of comprehension. Threshold exceedances are shown in grey. For ease of illustration on the regular grid, locations of observations with x-coordinate $\le 120^\circ$ and/or  y-coordinate $\le 20$ days, where $(120^\circ,20$days) corresponds to the node location nearest the origin) have been shifted by $+360$. For ease of use of the irregular grid, the triangulation is extended by repeating node locations $\pm 360$ in each covariate dimension.}
    \label{fig:2cov_grid}
\end{figure}

Node placement in 2-D presents a greater challenge than in 1-D. We adopt two approaches to node placement. The first approach (henceforth ``regular grid'') is defined in terms of $D$ ($=2$ here) sets of marginal node locations (on $[0,360)$ here), used together to produce a regular lattice on $\mathcal{D}=[0,360) \times [0,360)$. The first node location in each set is also used to locate an addition ``wrapping node'', located at the location of the first node plus 360 degrees or days, to accommodate periodicity on $\mathcal{D}$. The sets of marginal nodes are then used to create a regular rectangular grid of nodes on $\mathcal{D}$. Extra nodes are then added at the centres of each rectangle created, producing a (periodic) triangulation of $\mathcal{D}$. For a 2-D representation with $K_1$ nodes (or bins) directionally, and $K_2$ seasonally, the resulting total number $K$ of nodes is $K=2 K_1 K_2$, and the corresponding number of triangular bins is $M=4 K_1 K_2$. An example of a $3 \times 2$ regular grid (with three marginal nodes in direction and two in season, and a total of 12 nodes and 24 bins) is shown in the left panel Figure~\ref{fig:2cov_grid}. We refer to such a representation as a ``$K_1 \times K_2$'' grid. The intuitive approach explained in Section~\ref{sec:node} is again used to specify each set of marginal nodes.
The regular grid is attractive because it is specified in terms of sets of marginal nodes which are relatively straightforward to locate. 

The second approach to node specification (henceforth ``irregular grid'') permits ``freehand'' node location on $\mathcal{D}$. This approach is obviously more flexible, and potentially more parsimonious than the regular grid, but requires judicious choice of node locations in $D$ dimensions; this specification may be relatively straightforward in 2-D, but in general may prove challenging. Once the irregular node locations are specified, we create ``wrapping nodes'' with co-ordinates shifted by $\pm360$ (degrees or days) relative to those of the specified nodes in each covariate dimension. We then triangulate $\mathcal{D}$ using the specified and wrapping nodes. The right panel of Figure~\ref{fig:2cov_grid} illustrated an irregular grid with 6 specified nodes and 12 triangular bins. 

\begin{figure}
    \centering
    \includegraphics[scale=0.7]{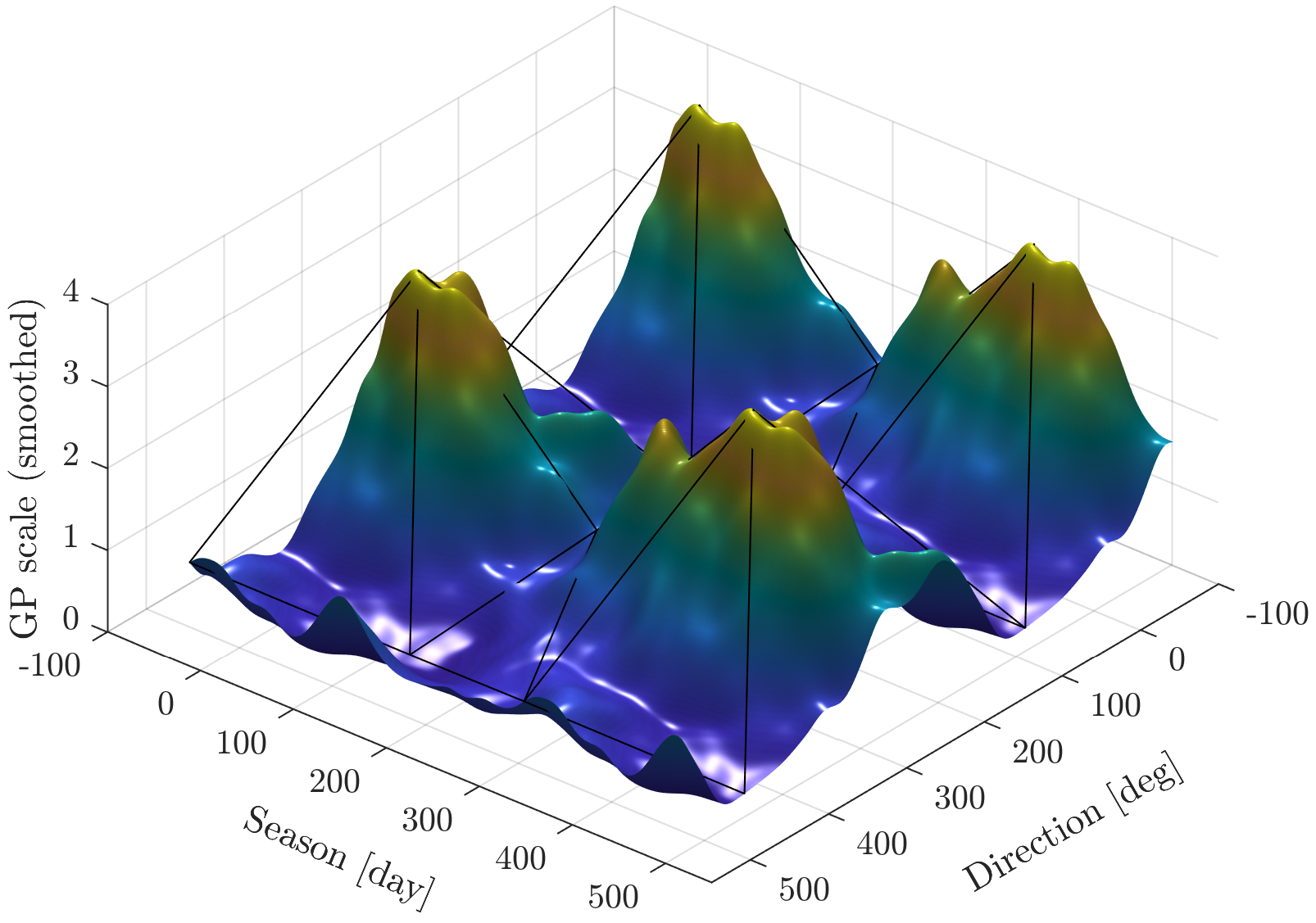}
    \captionit{Gaussian-kernel-smoothed initial local GP scale estimates. Grid lines between nodes (black lines) visualise a piecewise-linear scale representation. Reasonable agreement between actual value of scale, and that suggested by the piecewise linear triangulation, suggests good node placement. An extended covariate domain is used to aid interpretation.}
    \label{fig:2cov_init_scale}
\end{figure}

As for 1-D covariates, irregular nodes in $D$ dimensions should be placed where the local gradient of a model parameter with covariate is expected to change, i.e. at local turning points. We select node locations by inspection, again using initial local estimates of GP scale to guide node placement. An example is shown in Figure~\ref{fig:2cov_init_scale} (corresponding to a 2-D extension of Figure~\ref{fig:1cov_voronoi}). The surface shows initial local GP scale estimates from a local stationary GP fit to the $C=50$ nearest neighbours any location on $\mathcal{D}$, Gaussian-kernel-smoothed with common bandwidth $w=10$ degrees or days. Black grid lines emanating from nodes represent the irregular grid triangulation, and a piecewise-linear representation for GP scale. Node placement can be adjusted manually until reasonable agreement is obtained between actual value of scale, and that suggested by the piecewise linear triangulation.

We assess the predictive performance of non-stationary PPL extreme value models on 2-D covariate domains using different choices of regular and irregular grids. For regular grids, we consider $K_1=2$, 3, or 4 marginal directional nodes and $K_2=2$ or 3 marginal seasonal nodes. We consider irregular grids with $K=4$, 6, 8, 10, 12 and 14 nodes. In the present examples, node sets for irregular grids are nested, in the sense that nodes for a $K$-node grid are also nodes for the $K'$-node grid when $K \le K'$.  

\begin{figure}
    \centering
    \includegraphics[scale=0.5]{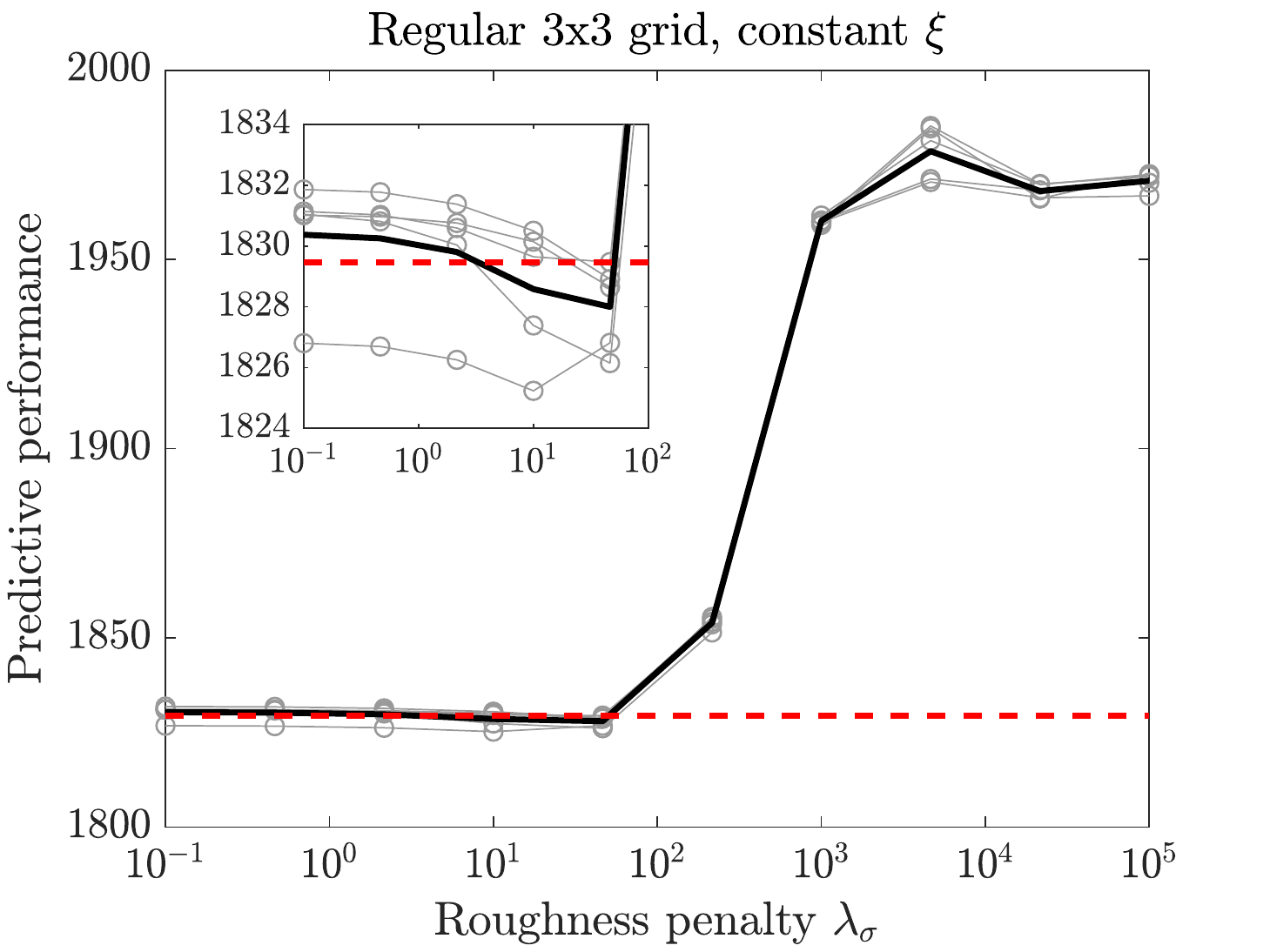}
    \captionit{Illustrative results for estimation of optimal roughness coefficient for a $3\times 3$ Case A model. Grey lines show negative log predictive likelihood $\mathcal{P}$on roughness coefficient $\lambda_{\sigma}$, for each of $R=5$ replicate analyses. The solid black line is the mean $\bar{\mathcal{P}}$ over the $R=5$ replicates. The dashed red line indicates the `accepted' value of predictive performance. Inset panel gives a magnified view near the minimum of $\bar{\mathcal{P}}$.}
    \label{fig:2cov_crossval}
\end{figure}

The cross-validation procedure used to estimate optimal values of roughness coefficients is the equivalent to that discussed in Section~\ref{sec:App:1D} for 1-D covariates, adapted to accommodate the appropriate number of roughness coefficients. Figure~\ref{fig:2cov_crossval} shows an example for a (regular) $3 \times 3$ Case A model, involving the estimation of a single roughness parameter $\lambda_{\sigma}$ for GP scale in both direction and season. Grey lines show negative log predictive likelihood ${\mathcal{P}}$ on roughness penalty $\lambda_{\sigma}$, for each of $R=5$ replicate analyses, and the solid black line represents the mean $\bar{\mathcal{P}}$ over the $R$ replicates. As for the 1-D case, the dashed red line shows the ``accepted'' value of predictive performance. The inset panel shows a magnified view in the vicinity of the minimum of $\bar{\mathcal{P}}$. The optimal roughness coefficient is found to be $\lambda_{\sigma}^*=10^{1.67}$ in this case.

Predictive performance for different regular and irregular grids is summarised in Tables \ref{tab:CV_reg} and \ref{tab:CV_irreg}. As for 1-D covariates, we find that more complex models sometimes show inferior performance corresponding to overly-heavy roughness penalisation (and hence over-fitting when not sufficiently penalised) to avoid problems of infinite negative log predictive likelihood during cross-validatory assessment. For the regular grids, allowing separate penalties for $\sigma$ in each covariate dimension (Case C) did not significantly improve the predictive performance compared to having a common penalty for $\sigma$ in both covariate dimensions (Case A). Therefore, for the irregular grids, only Cases A and B were considered. Allowing shape to vary with covariates (Case B) led to improved performance for some grids, but worse performance for others. In general the difference in performance was less than the uncertainty range in the estimated performance. Moreover, for most grids, the optimal penalties for the shape parameter force shape to be effectively constant with covariate.
\begin{table}
\centering
	\begin{tabular}{l | l | c c c c c c}
	& & \multicolumn{6}{c}{Numbers of marginal nodes, $(K_1 \times K_2)$}\\
	Case & Variable & $(2 \times 2)$ & $(3 \times 2)$ & $(4 \times 2)$ & $(2 \times 3)$ & $(3 \times 3)$ & $(4 \times 3)$\\
	\hline
	\multirow{2}{*}{A} & Optimal performance & 1837 (2.8) & 1828 (2.9) & 1839 (0.9) & 1830 (3.1) & 1828 (0.8) & 1863 (2.5)\\
    & $\log_{10}\left(\lambda_{\sigma}^*\right)$ & 1.67 & 1.67 & 2.33 & 1.67 & 1.67 & 2.33\\
    \hline
    \multirow{3}{*}{B} & Optimal performance & 1833 (1.5) & 1832 (3.2) & 1824 (7.6) & 1836 (3.6) & 1824 (2.3) & 1964 (3.7)\\
    & $\log_{10}\left(\lambda_{\sigma}^*\right)$ & 0.33 & 1.67 & 1.67 & 1.67 & 1.67 & 3\\
    & $\log_{10}\left(\lambda_{\xi}^*\right)$ & 2.33 & 5 & 3 & 3 & 2.33 & 3\\
    \hline
    \multirow{3}{*}{C} & Optimal performance & 1838 (1.3) & 1834 (2.3) & 1833 (2.8) & 1829 (1.7) & 1830 (3.0) & 1840 (2.7)\\
    & $\log_{10}\left(\lambda_{\sigma_1}^*\right)$ & 2.33 & 1.67 & 1.67 & 1.67 & 1.67 & 1.67\\
    & $\log_{10}\left(\lambda_{\sigma_2}^*\right)$ & 2.33 & 1 & 2.33 & 1.67 & 1.67 & 2.33
	\end{tabular}
	\captionit{Optimal predictive performance, $\bar{\mathcal{P}}(\boldsymbol{\lambda}^*)$, and optimal penalties, $\boldsymbol{\lambda}^*$, for 2-D regular grid models. Numbers in brackets are approximate uncertainties from jackknife analysis.}
	\label{tab:CV_reg}
\end{table}
\begin{table}
\centering
	\begin{tabular}{l | l | c c c c c c}
	& & \multicolumn{6}{c}{Number of nodes, $K$}\\
	Case & Variable & 4 & 6 & 8 & 10 & 12 & 14\\
	\hline
	\multirow{2}{*}{A} & Optimal performance & 1836 (1.5) & 1816 (2.3) & 1815 (2.1) & 1814 (4.6) & 1814 (3.6) & 1813 (4.8)\\
    & $\log_{10}\left(\lambda_{\sigma}^*\right)$ & 2.33 & 2.33 & 1.67 & 2.33 & 1.67 & 1.67\\
    \hline
    \multirow{3}{*}{B} & Optimal performance & 1834 (1.4) & 1816 (2.6) & 1815 (3.1) & 1809 (3.1) & 1809 (2.8) & 1815 (2.2) \\
    & $\log_{10}\left(\lambda_{\sigma}^*\right)$ & 1.67 & 2.33 & -0.33 & 1.67 & 1.67 & 1.0\\
    & $\log_{10}\left(\lambda_{\xi}^*\right)$ & 2.33 & 2.33 & 5.0 & 5.0 & 4.33 & 5.0\\
	\end{tabular}
	\captionit{Optimal predictive performance, $\bar{\mathcal{P}}(\boldsymbol{\lambda}^*)$, and optimal penalties, $\boldsymbol{\lambda}^*$, for 2-D irregular grid models. Numbers in brackets are approximate uncertainties from jackknife analysis.}
	\label{tab:CV_irreg}
\end{table}

Irregular grid models with six or more nodes give better performance than regular grid models. Even the simplest 4-node irregular grid model gives comparable performance to the regular grids; this feature is attributed to careful irregular grid node placement, allowing more parsimonious description of parameter variation. Although there is some improvement in the predictive performance for irregular grids with additional nodes, these improvements are small compared with jackknife performance uncertainties.

\begin{figure}
    \centering
    \includegraphics[scale=0.55]{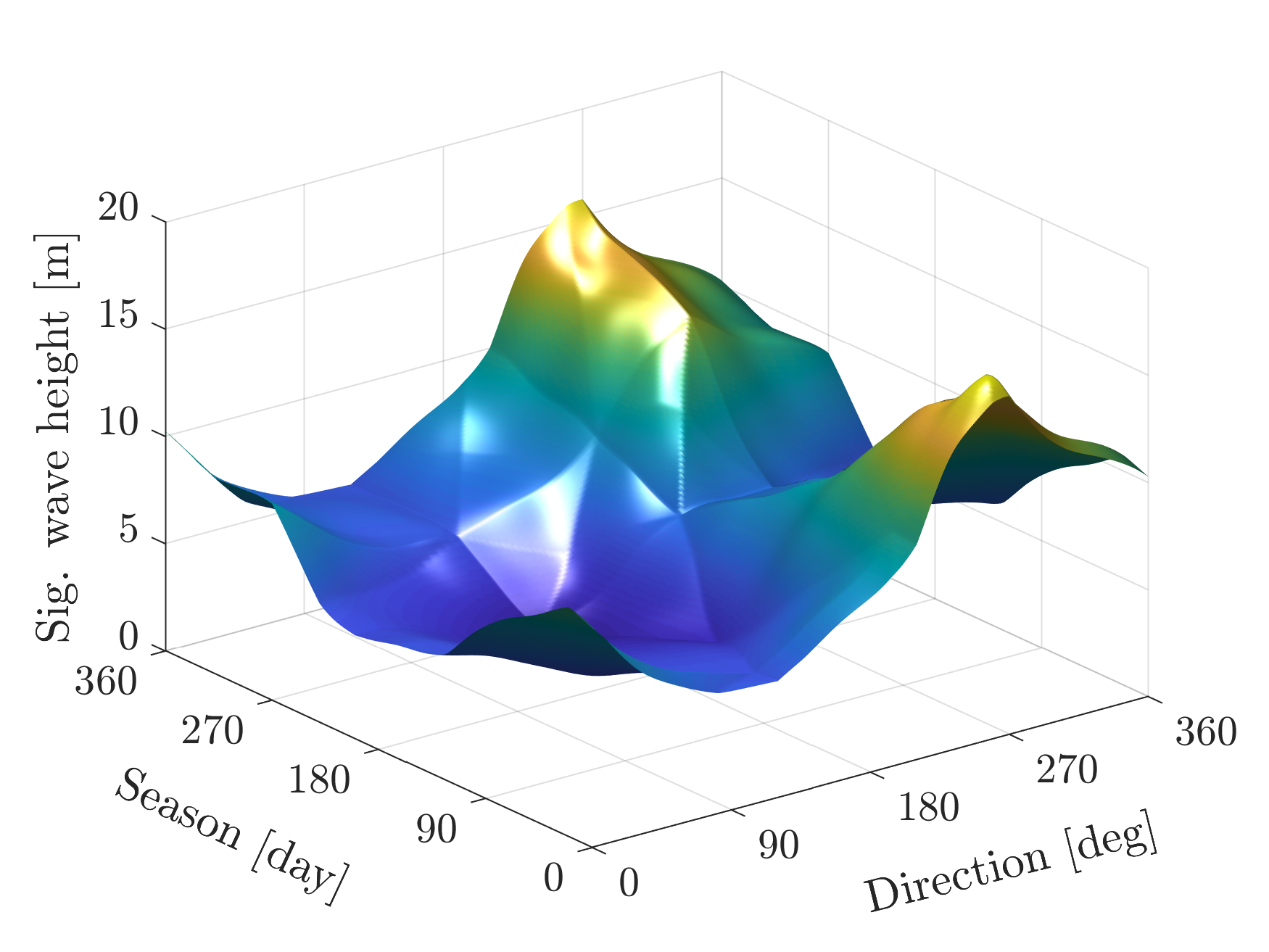}
    \includegraphics[scale=0.55]{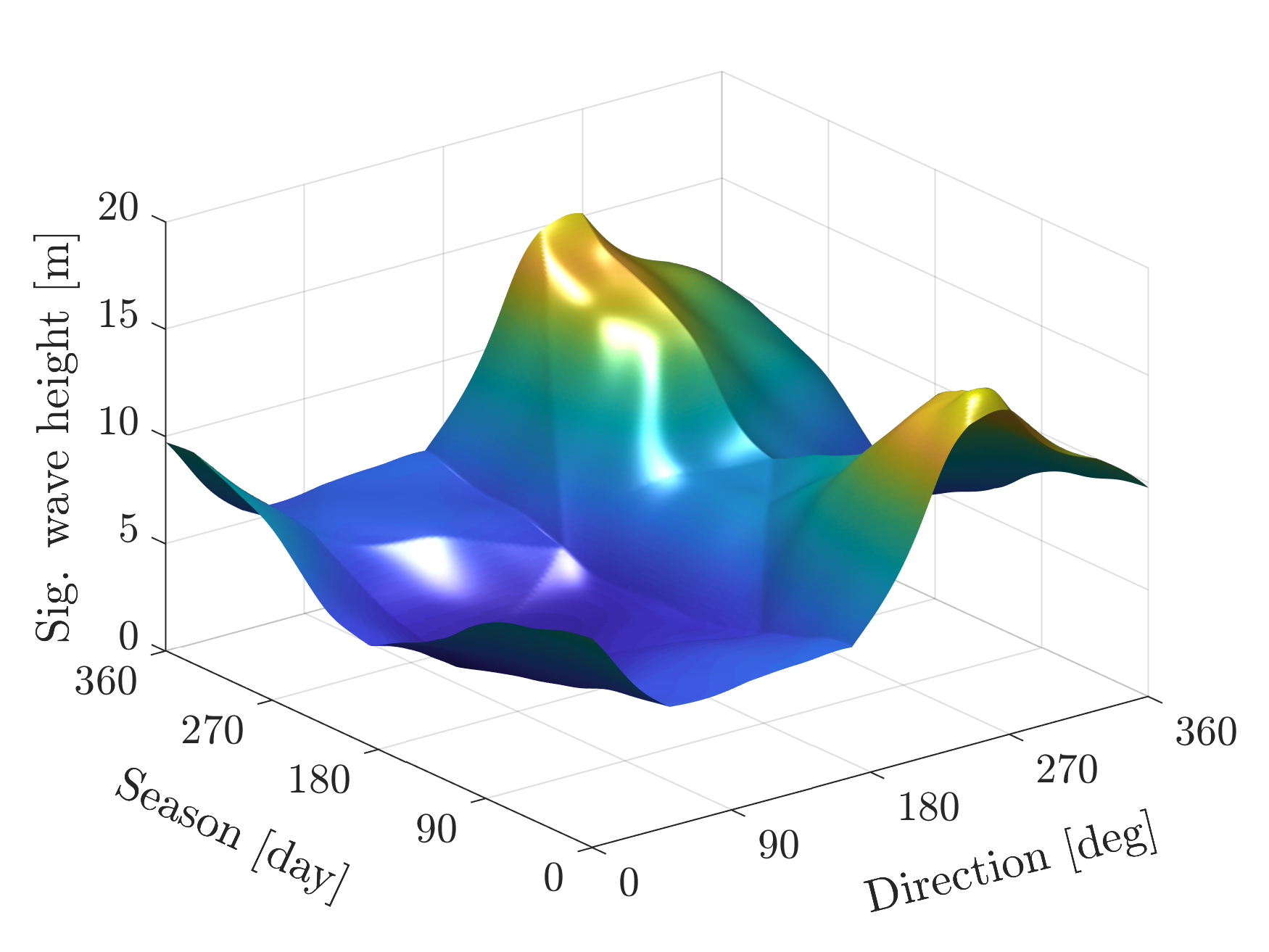}
    \captionit{Directional-seasonal conditional quantile of $H_S$ for non-exceedance probability 0.99, for regular $3\times 3$ and irregular 6-node Case A models (left and right respectively).}
    \label{fig:2cov_quant}
\end{figure}

Figure~\ref{fig:2cov_quant} shows directional-seasonal quantiles under the regular $3 \times 3$ and irregular 6-node Case A models, corresponding to non-exceedance probability 0.99. 
The effects of both the piecewise-linear grid for $\sigma(\mathbf{x})$, and the nonlinear non-parametric threshold estimate are visible. The irregular grid provides a more adequate description of the sharp transition in $H_S$ around $210^\circ$, due to judicious node placement. However, quantile estimates are generally similar, indicating that inferences are not strongly dependent on grid choice.

\begin{figure}
    \centering
    \includegraphics[scale=0.5]{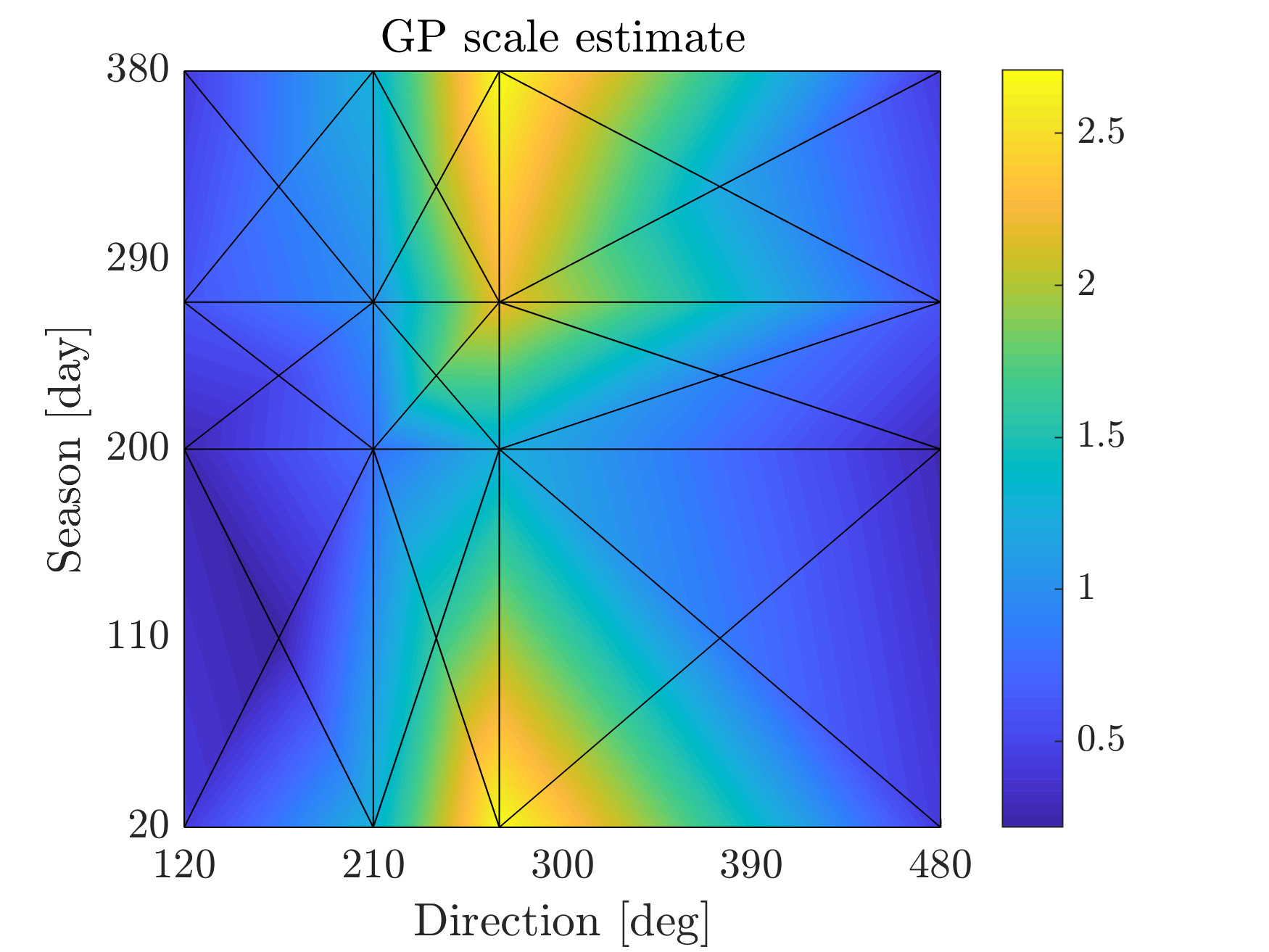}
    \includegraphics[scale=0.5]{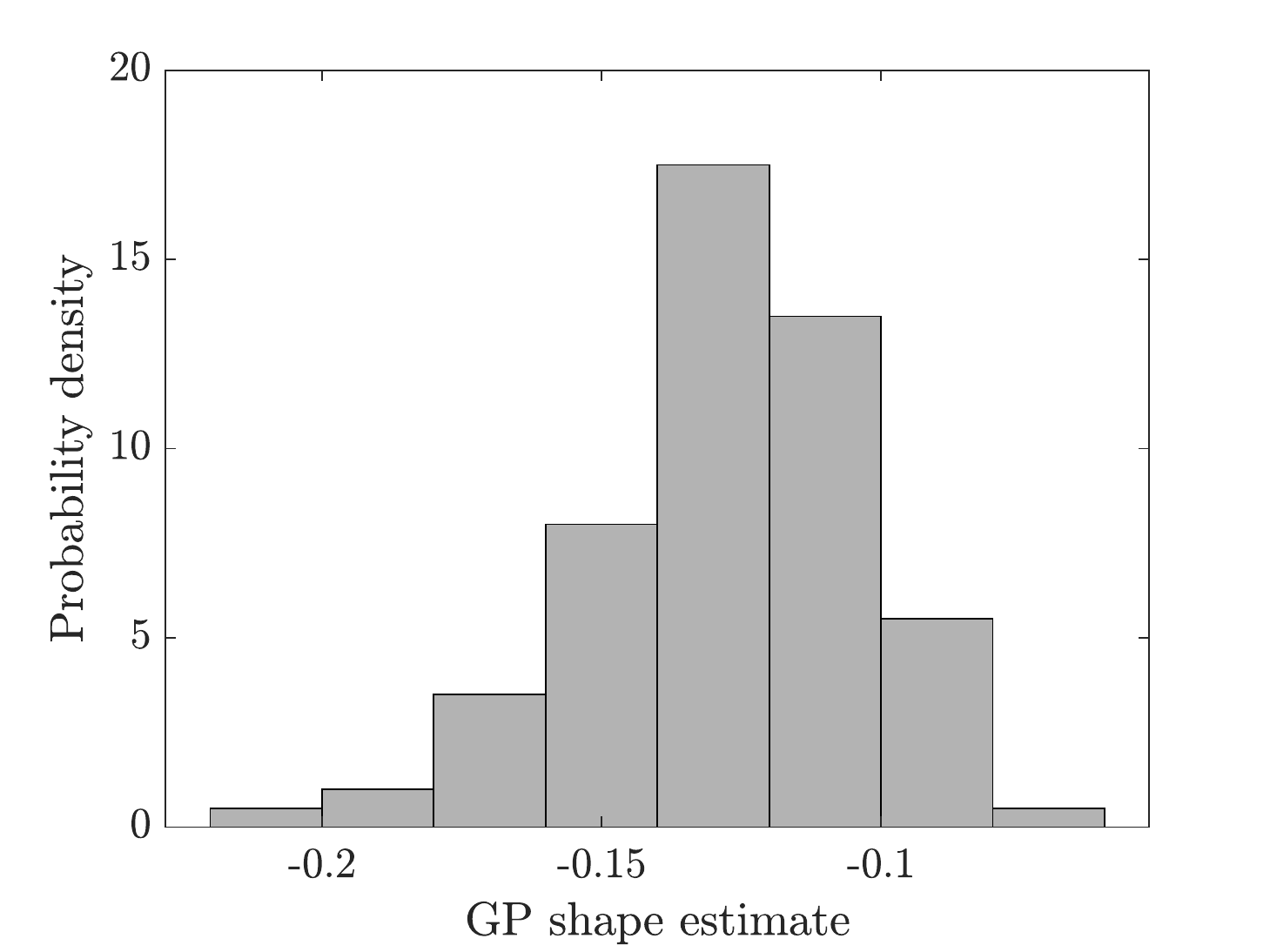}
    \captionit{Estimates of GP scale and shape for the regular $3\times 3$ Case A model. Left: mean directional-seasonal scale estimate. Right: distribution of (stationary) shape estimates.}
    \label{fig:2cov_param}
\end{figure}

\begin{figure}
    \centering
    \includegraphics[scale=0.5]{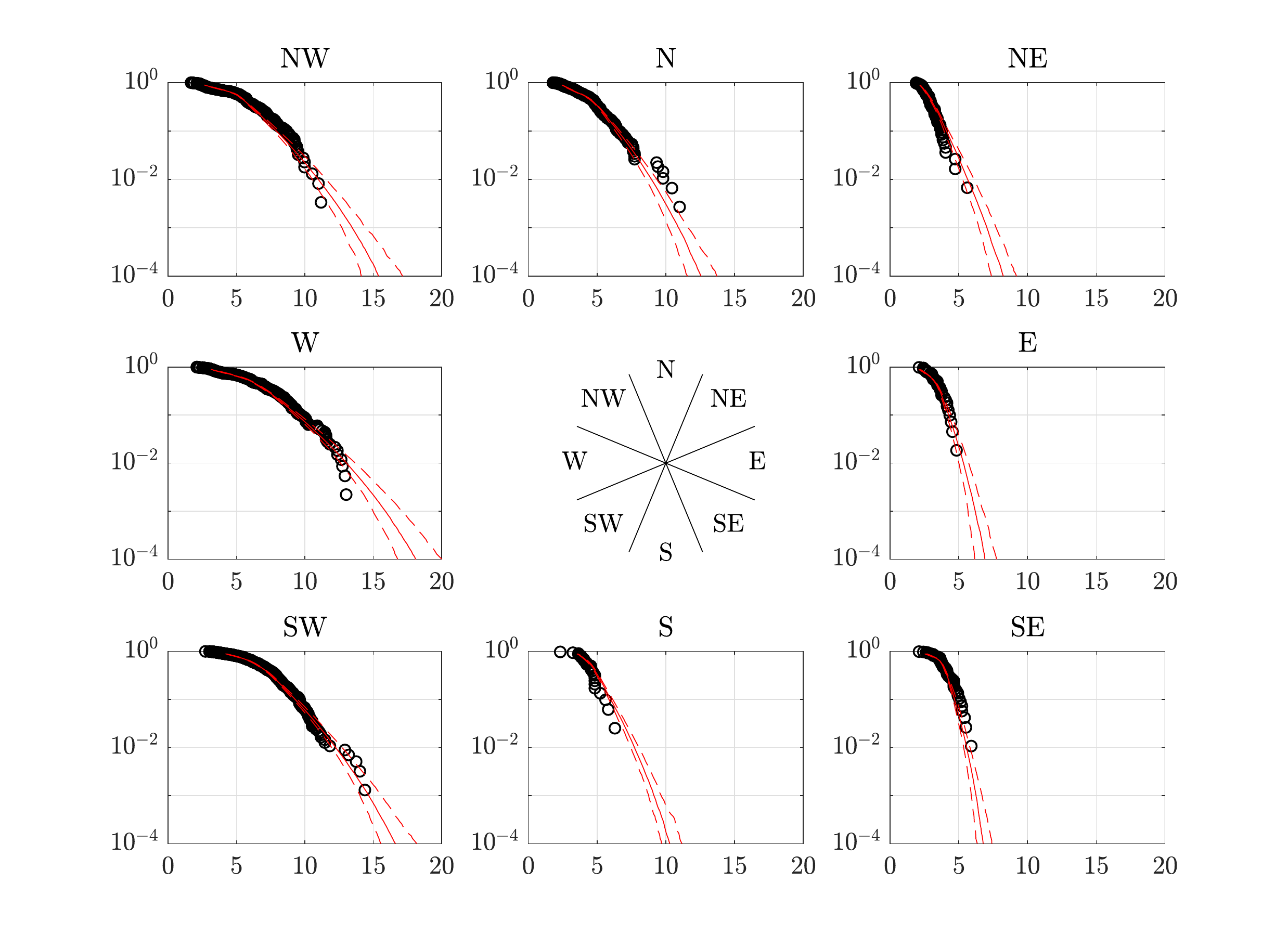}
    \captionit{Tail plots per directional octant from the irregular $6$-node directional-seasonal Case A model. Abscissa is significant wave height in metres. Ordinate is exceedance probability. Empirical estimates in black, and model-based estimates in red. Dashed lines represent 95\% uncertainty bands based on 100 bootstrap trials.}
    \label{fig:2cov_exceed_dir}
\end{figure}

Parameter estimates for the regular $3 \times 3$ Case A model are shown in Figure~\ref{fig:2cov_param}. The left panel shows the mean scale estimate of inferences using each of 100 bootstrap resamples. The right panel shows the corresponding distribution of stationary shape estimate. The general features of the left panel reflect those of the corresponding covariate density and threshold in Figure~\ref{fig:2cov_pdf_thresh}. Values of parameter estimates are further similar to those for the corresponding 1-D directional model illustrated in Figure~\ref{fig:1cov_dir_param}.
An illustrative directional tail plot for the irregular 6-node directional-seasonal Case A model is given in Figure~\ref{fig:2cov_exceed_dir}. Agreement between observations and the fitted model is good.

\section{Discussion and conclusions}  \label{sec:Dsc}
Capturing covariate effects in ocean environmental extreme value models 
can be important for optimising the design of an asymmetric structure with respect to direction, or assessing risk associated with an operation in a given season. It can also potentially improve inference for extreme quantiles. The level of sophistication required to describe covariate effects depends on the characteristics of the sample. We expect that a relatively simple penalised piecewise-linear (PPL) representation for the functional forms of extreme value model parameters will often be sufficient. In this paper, we demonstrate that the PPL generalised Pareto (GP) model for peaks over threshold of significant wave height provides good characterisation of directional and seasonal effects at low computational cost. Roughness penalisation, regulated using a cross-validation scheme, ensures that parsimonious models are estimated. Uncertainties in parameter and return value estimates are quantified using bootstrap re-sampling. MATLAB software for the PPL model is available at \url{https://github.com/edmackay/PPL-model}, along with a user-guide and the data underpinning the current study.

Relative to more sophisticated covariate representations (e.g. \citealt{Wod11}, \citealt{ZnnEA19a}), the PPL model is conceptually simple to understand, and facilitates computationally-simple inference, both of which we consider to be attractive features for the metocean practitioner. Relative to the simpler penalised piecewise-constant (PPC) approach of \cite{RssEA19}, the additional flexibility of the PPL model allows a physically more plausible representation of continuous parameter variation on the covariate domain. Better use of the sample is made, with parameter values at each PPL node informed by observations in all bins adjacent to that node. In contrast, the PPC model is only penalised for the total variance of parameter values over the bins, ignoring bin proximity. 

In the current work, as outlined in Section~\ref{sec:MdlFrm:Jnt} and Equation~\ref{eq:XY_taildensity}, we are interested in estimating the joint distribution $f_{\mathbf{X},Y}(\mathbf{x},y)$ for $y>u(\mathbf{x})$. This requires models for the extreme value threshold ($u(\mathbf{x})$, in order to extract threshold exceedances, and estimate their density $f_{\text{GP}}(y|u(\mathbf{x}), \sigma(\mathbf{x}), \xi(\mathbf{x}))$) and the density $f_{\mathbf{X}}(\mathbf{x})$ of covariates. Unlike the GP model for threshold exceedances, estimation for both $u(\mathbf{x})$ and $f_{\mathbf{X}}(\mathbf{x})$ utilises the full sample. It is therefore reasonable to seek more detailed descriptions, using flexible local estimators, for covariate density and threshold, than for the density of threshold exceedances. In this sense, the current hierarchical approach imposes increasing regularity on non-stationary effects as the statistical efficiency of the estimation decreases, incorporating kernel density estimation for covariate density and local quantile estimation for threshold, but a more rigid piecewise-linear model for GP scale, and either a constant or piecewise-linear model for GP shape. However, as implemented here, uncertainties from the estimation of covariate density and threshold are not propagated into the extreme value analysis; estimates of extreme value models here are therefore conditional on the estimated extreme value threshold. Extreme quantile estimates are further conditional on the estimated covariate density. The effect of threshold uncertainty on return values can be large; in the current work, an appropriate threshold non-exceedance probability is selected to ensure that inferences are stable with respect to small variation in this choice. More complex approaches (e.g. \citealt{RndEA15a}) seek to estimate covariate density, threshold and extreme value parameters in a single inference, at the cost of increased model complexity and computational burden. Due to the relative sparsity of information for the GP shape parameter, its roughness coefficient is often relatively large, implying a smooth variation of the estimate on the covariate domain. Nevertheless, models admitting non-constant shape perform slightly better in general compared to those with constant shape: allowing a little variation in shape improves performance.

Choice of the number and locations of nodes is key to the success of the PPL methodology. Diagnostic tools can aid these choices. In 2-D, it appears that judiciously-chosen ``irregular'' nodes provide the best predictive performance. Nevertheless, models exploiting nodes located on a ``regular'' rectangular grid, specified in terms of nodes for marginal components, also perform relatively well. In the current work, we have assumed that the same node locations are appropriate for GP scale and shape estimation; clearly this may not always be the case, and we might expect that fewer nodes might be sufficient to describe shape parameter variation. Nevertheless, the optimal choice of roughness coefficients goes some way to providing balanced parameterisations across covariate dimensions and GP parameters. In general, extreme value modelling using Bayesian inference, exploiting reversible jump or similar methodologies in 1-D (e.g. \citealt{ZnnEA19a}) and 2-D (e.g. \citealt{Jnt21Saes}) provide approaches to estimate the number and location of nodes automatically as part of the inference, at increased computational cost for a more complex model formulation. Experience also suggests that the Bayesian paradigm which permits, via stochastic sampling schemes (e.g. the Metropolis-Hastings or mMALA algorithms), an exploration of the full posterior distribution of the parameters, is more reliable than the frequentist approach. The latter relies on accurate identification of the global mode of a high-dimensional and often highly multi-modal surface; in many cases it is difficult to either identify the mode or to verify that the mode is indeed global. 

In the current analysis, the number of nodes is kept to a reasonably small value, since we are motivated by the idea of providing the conceptually simplest but useful extreme value model for the sample. However, the statistical literature on penalised B-splines (e.g. \citealt{ElrMrx10}) recommends specification of a large number of spline nodes (or knots), with roughness penalisation subsequently imposing the optimal smoothness on the solution, reducing the effective number of degrees of freedom in the covariate representation. In our case, with small numbers of nodes, penalisation plays a less important role in general, but is nevertheless important when the number of nodes is over-specified. Moreover, as a relatively simple optimisation method has been used in the present work, using a small number of nodes makes finding an optimal set of parameter estimates feasible, without excessive computational effort. 

Further work might include a systematic comparison of PPL performance relative to its peers, and perhaps also relative to methods involving transformation of data to standard scales prior to extreme value analysis (e.g. \citealt{EstTwn12}). The current work has addressed periodic covariates only; inference for non-periodic covariates is in some senses simpler. However, in other applications (e.g. \citealt{Mackay2020assessment}) a non-periodic covariate may itself become extreme, requiring joint extreme value modelling of ``covariate'' and response.
    
\section*{Acknowledgement}
This work was funded by the UK EPSRC Supergen Offshore Renewable Energy Hub, project EP/S000747/1 on ‘‘Improved Models for Multivariate Metocean Extremes (IMEX)’’.

\bibliography{PPLpaper}

\begin{thebibliography}{}

\bibitem[Anderson et~al., 2001]{AndCrt01}
Anderson, C., Carter, D., and Cotton, P. (2001).
\newblock {\em Wave climate variability and impact on offshore design
  extremes}.
\newblock Report commissioned from the \textsc{U}niversity of
  \textsc{S}heffield and \textsc{S}atellite \textsc{O}bserving \textsc{S}ystems
  for \textsc{S}hell \textsc{I}nternational.

\bibitem[Carter and Challenor, 1981]{CrtChl81}
Carter, D. J.~T. and Challenor, P.~G. (1981).
\newblock Estimating return values of environmental parameters.
\newblock {\em Quarterly Journal of the Royal Meteorological Society}, 107:259.

\bibitem[Chavez-Demoulin and Davison, 2005]{ChvDvs05}
Chavez-Demoulin, V. and Davison, A. (2005).
\newblock Generalized additive modelling of sample extremes.
\newblock {\em Journal of the Royal Statistical Society C: Applied Statistics},
  54:207--222.

\bibitem[Cleveland, 1979]{Clv79}
Cleveland, W.~S. (1979).
\newblock Robust locally weighted regression and smoothing scatterplots.
\newblock {\em Journal of the American Statistical Association}, 74:829--836.

\bibitem[Coles and Walshaw, 1994]{ClsWls94}
Coles, S. and Walshaw, D. (1994).
\newblock Directional modelling of extreme wind speeds.
\newblock {\em Applied Statistics}, 43:139--157.

\bibitem[Davison and Smith, 1990]{DvsSmt90}
Davison, A. and Smith, R.~L. (1990).
\newblock Models for exceedances over high thresholds.
\newblock {\em Journal of the Royal Statistical Society B}, 52:393.

\bibitem[Eastoe and Tawn, 2012]{EstTwn12}
Eastoe, E. and Tawn, J. (2012).
\newblock Modelling non-stationary extremes with application to surface level
  ozone.
\newblock {\em Journal of the Royal Statistical Society C: Applied Statistics},
  58:25--45.

\bibitem[Eastoe, 2019]{Eastoe2019}
Eastoe, E.~F. (2019).
\newblock {Nonstationarity in peaks-over-threshold river flows: A regional
  random effects model}.
\newblock {\em Environmetrics}, 30(5):1--18.

\bibitem[Eilers and Marx, 2010]{ElrMrx10}
Eilers, P. H.~C. and Marx, B.~D. (2010).
\newblock Splines, knots and penalties.
\newblock {\em Wiley Interscience Reviews: Computational Statistics},
  {2}:{637--653}.

\bibitem[Ewans and Jonathan, 2008]{EwnJnt08}
Ewans, K.~C. and Jonathan, P. (2008).
\newblock The effect of directionality on northern \textsc{N}orth \textsc{S}ea
  extreme wave design criteria.
\newblock {\em Journal of Offshore Mechanics and Arctic Engineering},
  130:041604:1--041604:8.

\bibitem[Fawcett and Walshaw, 2007]{FwcWls07}
Fawcett, L. and Walshaw, D. (2007).
\newblock Improved estimation for temporally clustered extremes.
\newblock {\em Environmetrics}, 18:173--188.

\bibitem[Guerrero et~al., 2021]{Guerrero2021}
Guerrero, M.~B., Huser, R., and Ombao, H. (2021).
\newblock {Conex-Connect: Learning Patterns in Extremal Brain Connectivity From
  Multi-Channel EEG Data}.
\newblock {\em Pre-print}.

\bibitem[Haver, 1987]{Hvr87}
Haver, S. (1987).
\newblock On the joint distribution of heights and periods of sea waves.
\newblock {\em Ocean Engineering}, 14:359--376.

\bibitem[Jonathan, 2021]{Jnt21Saes}
Jonathan, P. (2021).
\newblock {A non-parametric representation for multidimensional covariates in
  an extreme value model}.
\newblock {\em Gwerddon}, 33:68--84.

\bibitem[Jonathan et~al., 2010]{JntFlnEwn10}
Jonathan, P., Flynn, J., and Ewans, K.~C. (2010).
\newblock Joint modelling of wave spectral parameters for extreme sea states.
\newblock {\em Ocean Engineering}, 37:1070--1080.

\bibitem[Konzen et~al., 2021]{KnzEA21}
Konzen, E., Neves, C., and Jonathan, P. (2021).
\newblock Modeling nonstationary extremes of storm severity: Comparing
  parametric and semiparametric inference.
\newblock {\em Environmetrics}, 32:e2667.

\bibitem[Mackay et~al., 2010]{Mackay2010}
Mackay, E., Challenor, P.~G., and Bahaj, A. B.~S. (2010).
\newblock {On the use of discrete seasonal and directional models for the
  estimation of extreme wave conditions}.
\newblock {\em Ocean Engineering}, 37:425--442.

\bibitem[Mackay and Jonathan, 2020a]{Mackay2020assessment}
Mackay, E. and Jonathan, P. (2020a).
\newblock Assessment of return value estimates from stationary and
  non-stationary extreme value models.
\newblock {\em Ocean Engineering}, 207.

\bibitem[Mackay and Jonathan, 2020b]{Mackay2020omae}
Mackay, E. and Jonathan, P. (2020b).
\newblock {Estimation of environmental contours using a block resampling
  method}.
\newblock {\em Proceedings of the International Conference on Offshore
  Mechanics and Arctic Engineering - OMAE2020}, 2A-2020.

\bibitem[Mackay and Jonathan, 2021]{Mackay2021sampling}
Mackay, E. and Jonathan, P. (2021).
\newblock {Sampling properties and empirical estimates of extreme events}.
\newblock {\em Ocean Engineering}, 239:109791.

\bibitem[Mendez et~al., 2008]{MndEA08}
Mendez, F.~J., Menendez, M., Luceno, A., Medina, R., and Graham, N.~E. (2008).
\newblock Seasonality and duration in extreme value distributions of
  significant wave height.
\newblock {\em Ocean Engineering}, 35:131--138.

\bibitem[Northrop et~al., 2017]{NrtEA15}
Northrop, P., Attalides, N., and Jonathan, P. (2017).
\newblock Cross-validatory extreme value threshold selection and uncertainty
  with application to ocean storm severity.
\newblock {\em Journal of the Royal Statistical Society C: Applied Statistics},
  66:93--120.

\bibitem[Northrop and Jonathan, 2011]{NrtJnt11}
Northrop, P. and Jonathan, P. (2011).
\newblock Threshold modelling of spatially-dependent non-stationary extremes
  with application to hurricane-induced wave heights.
\newblock {\em Environmetrics}, 22:799--809.

\bibitem[Randell et~al., 2015]{RndEA15}
Randell, D., Feld, G., Ewans, K., and Jonathan, P. (2015).
\newblock Distributions of return values for ocean wave characteristics in the
  {S}outh {C}hina {S}ea using directional-seasonal extreme value analysis.
\newblock {\em Environmetrics}, 26:442--450.

\bibitem[Randell et~al., 2016]{RndEA15a}
Randell, D., Turnbull, K., Ewans, K., and Jonathan, P. (2016).
\newblock Bayesian inference for non-stationary marginal extremes.
\newblock {\em Environmetrics}, 27:439--450.

\bibitem[Reistad et~al., 2011]{RstEA11}
Reistad, M., Breivik, O., Haakenstad, H., Aarnes, O.~J., Furevik, B.~R., and
  Bidlot, J.-R. (2011).
\newblock A high-resolution hindcast of wind and waves for the {N}orth {S}ea,
  the {N}orwegian {S}ea, and the {B}arents {S}ea.
\newblock {\em J. Geophys. Res.}, 116:1--18.

\bibitem[Robinson and Tawn, 1997]{RbsTwn97}
Robinson, M.~E. and Tawn, J.~A. (1997).
\newblock Statistics for extreme sea currents.
\newblock {\em Applied Statistics}, 46:183--205.

\bibitem[Ross et~al., 2020]{RssEA19}
Ross, E., Astrup, O.~C., Bitner-Gregersen, E., Bunn, N., Feld, G., Gouldby, B.,
  Huseby, A., Liu, Y., Randell, D., Vanem, E., and Jonathan, P. (2020).
\newblock {On environmental contours for marine and coastal design}.
\newblock {\em Ocean Engineering}, 195:106194.

\bibitem[Ross et~al., 2018]{Ross2018}
Ross, E., Sam, S., Randell, D., Feld, G., and Jonathan, P. (2018).
\newblock {Estimating surge in extreme North Sea storms}.
\newblock {\em Ocean Engineering}, 154(January):430--444.

\bibitem[Scotto and Guedes-Soares, 2000]{SctGds00}
Scotto, M. and Guedes-Soares, C. (2000).
\newblock Modelling the long-term time series of significant wave height with
  non-linear threshold models.
\newblock {\em Coastal Engineering}, 40:313--327.

\bibitem[Sigauke and Bere, 2017]{Siguake2017}
Sigauke, C. and Bere, A. (2017).
\newblock {Modelling non-stationary time series using a peaks over threshold
  distribution with time varying covariates and threshold: An application to
  peak electricity demand}.
\newblock {\em Energy}, 119:152--166.

\bibitem[Towe et~al., 2017]{TowEA17}
Towe, R., Eastoe, E., Tawn, J., and Jonathan, P. (2017).
\newblock {Statistical downscaling for future extreme wave heights in the North
  Sea}.
\newblock {\em Annals of Applied Statistics}, 11:2375--2403.

\bibitem[Wood, 2011]{Wod11}
Wood, S.~N. (2011).
\newblock Fast stable restricted maximum likelihood and marginal likelihood
  estimation of semiparametric generalized linear models.
\newblock {\em Journal of the Royal Statistical Society. Series B: Statistical
  Methodology}, 73(1):3--36.

\bibitem[Yang et~al., 2016]{YngEA16}
Yang, L., Liu, S., Tsoka, S., and Papageorgiou, L.~G. (2016).
\newblock Mathematical programming for piecewise linear regression analysis.
\newblock {\em Expert Systems with Applications}, 44:156--167.

\bibitem[Zanini et~al., 2020]{ZnnEA19a}
Zanini, E., Eastoe, E., Jones, M., Randell, D., and Jonathan, P. (2020).
\newblock {Covariate representations for non-stationary extremes}.
\newblock {\em Environmetrics}, 31:e2624.

\end{thebibliography}

\end{document}